%% file: ms.tex
\documentclass[preprint2]{aastex}
\usepackage{natbib}
\usepackage{lscape}
\bibpunct{(}{)}{;}{a}{}{,}

\def\Min{${}^{\prime}$\llap{.}}
\def\Sec{${}^{\prime\prime}$\llap{.}}
\def\timesec{${}^{s}$\llap{.}}
\def\deg{${}^\circ$}
\def\min{${}^{\prime}$}
\def\sec{${}^{\prime\prime}$}

\shorttitle{X-ray properties of protostars in Orion}
\shortauthors{Prisinzano et al.}

\begin{document}


\title{X-ray properties of protostars in the  Orion Nebula}

\author{L. Prisinzano\altaffilmark{1}, 
	G. Micela\altaffilmark{1},
	E. Flaccomio\altaffilmark{1},
	J. R. Stauffer\altaffilmark{2},
	T. Megeath\altaffilmark{3},
	L. Rebull\altaffilmark{2},
	M. Robberto\altaffilmark{4},
	K. Smith\altaffilmark{4},
	E. D. Feigelson\altaffilmark{5},
	N. Grosso\altaffilmark{6} and
	S. Wolk\altaffilmark{7}}




	  
	  



\altaffiltext{1}{INAF - Osservatorio Astronomico di Palermo,
   		Piazza del Parlamento 1,  
		I-90134 Palermo, Italy, loredana@astropa.inaf.it}
\altaffiltext{2}{Spitzer Science Center, California Institute of Technology, 
	 1200 East  
	 California Boulevard, Pasadena, CA 91125}
\altaffiltext{3}{Department of Physics, University of Toledo, 2801 W. 
	Bancroft Ave, 
	 Toledo OH 43606}
\altaffiltext{4}{Space Telescope Science Institute, 3700 San Martin Drive, 
Baltimore,  MD 21218}
\altaffiltext{5}{Department of Astronomy and Astrophysics, Penn State
University,  
525	 Davey Lab, University Park, PA 16802, USA}
\altaffiltext{6}{Observatoire astronomique de Strasbourg, Universit\'e 
	Louis-Pasteur, 
	 CNRS, INSU, 11 rue de l'Universit\'e, 67000 Strasbourg, France}
\altaffiltext{7}{Harvard Smithsonian Center for Astrophysics, MS 65, 60 
	Garden Street, 
	  Cambridge, MA 02138}




\begin{abstract}
The origin and evolution of the X-ray emission in very young stellar objects (YSOs)
are not yet well understood because it is very hard to observe YSOs in the
protostellar phase.
We study the X-ray properties of Class\,0-I objects in the Orion Nebula
 Cluster (ONC) and compare them with those of the more evolved 
 Class\,II and III members. Using Chandra Orion Ultradeep Project (COUP) data,
   we study the X-ray properties of stars in different evolutionary 
   classes: 
  luminosities, hydrogen column densities N$_{\rm H}$, average 
  plasma temperatures and  time variability are  compared
     in order  to understand if the interaction between the circumstellar
  material and the central object can influence the X-ray
  emission. 
We have assembled the deepest and most complete photometric catalog 
of objects in the ONC region from the UV 
to 8\,$\mu$m using data from
 the HST Treasury Program, deep and almost simultaneous UBVI
and JHK 
images  taken, respectively,
with WFI@2.2m ESO and ISPI@4m CTIO telescopes, and   {\it  Spitzer}  IRAC imaging.
We select high probability candidate 
Class\,0-I
protostars, distinguishing between  those
having a spectral energy distribution  which rises from K up to 8\,$\mu$m
(Class\,0-Ia) from those  where the SED rises from K up to 4.5\,$\mu$m 
and decreasing afterwards (Class\,0-Ib).
In addition, we select a sample of ``bona fide"
Class\,II stars 
  and a set of Class\,III stars with IR emission
consistent with normal photospheres.
  
Our principal result is that  Class\,0-Ia objects are significantly less luminous 
in X-rays, both in the total and hard bands,
 than  
 the more evolved Class\,II stars with mass larger than 0.5\,M$_\odot$;
 these latter show X-ray luminosities similar to those of Class\,0-Ib stars.
This result supports the hypothesis that the onset of X-ray emission occurs at a very 
early stage of star formation and is in agreement with the result found in \citet{giar07}.
 Spectral properties of Class\,0-I stars are similar to those of the more
 evolved Class\,II and III objects, except for a larger absorption likely due to
 gas in the envelope or disk of  the protostellar objects. Our data
 suggest that the three different classes have similar X-ray temporal
 variability. 
\end{abstract}





\section{INTRODUCTION}
 Young stellar objects (YSOs)  exhibit an X-ray luminosity
significantly higher than similar stars in the main sequence phase. However,
it is still unclear at what stage of their formation process YSOs begin to emit
X-rays  and how this X-ray emission evolves 
with time \citep{fava03,feig07}. 

Evidence has accumulated over the past few years that among the 
YSOs, the X-ray luminosity
function (XLF) of Classical T Tauri stars (CTTS) is different from that of 
Weak T Tauri stars (WTTS).
Such a difference has been investigated by some 
because
the accretion and outflow processes, characteristic of the CTTS  
phase, involve the interaction of
ionized material with the star and disk magnetic field, and
these media play a 
fundamental role in the coronal X-ray emission of young stars.  
  Using the {\it Chandra} Orion Ultradeep Project (COUP)  data
 \citep{getm05a}, \citet{prei05b} find that the accreting stars,
 as discriminated using the Ca\,{\footnotesize II} 8542\AA\ IR triplet line,
  are less  X-ray luminous than
 non-accretors, while no differences are found for stars with $(K-L)$~
 IR excesses,  indicative of 
 the presence of a (possibly passive) disk,
 with respect to stars without IR excesses. 
 These results are in agreement with those found  in the previous work 
 in the  ONC by \citet{flac03b,flac03c} 
 who found a strong difference in the X-ray
 luminosity of accreting and non-accreting stars.
 
  In order to understand the evolution of the X-ray emission during the star
  formation process, \citet{prei05a} looked
 for a possible relation between X-ray emission and stellar age
  in the ONC. They found, in the  age range $\sim0.1-10$\,Myr, 
  a slow decay in the X-ray luminosity, L$_{\rm X}$. 
Because, as the star contracts, 
  the bolometric luminosity, L$_{\rm bol}$,
decreases faster than L$_{\rm X}$, 
  on average L$_{\rm X}$/L$_{\rm bol}$ and the X-ray surface flux, F$_{\rm X}$,
   increase during this age range.

  In this work, we want to investigate the onset and the evolution 
  of X-ray emission during the early phase of the star formation by
  statistically comparing the X-ray properties of the stars in different evolutionary phases
  rather than considering 
individual stellar ages, usually derived from theoretical
isochrones. We hope in this way to avoid some of the problems introduced
when using inferred ages (due to their often large random and
systematic uncertainties; see, e.g., \citeauthor{hill07} 2007).

 YSOs in their early phases  are characterized
 by IR excesses due to  the presence of a circumstellar envelope and/or a disk
 that contains warm dust. The first IR classification in Class\,I, II and III 
 sources, which is based on the IR spectral index, was proposed by 
 \citet{lada87}.  The discovery in the sub-mm of cold objects, younger than the
 IR sources, leads to the introduction of the Class\,0 sources \citep{andr93}. 
 Based on their mid-IR properties, YSOs
are usually classified as follows 
 \citep[e.g.][]{alle04,mege04,lada06,alle07}: 
 Class\,0 objects 
are protostars surrounded by the initial 
 infalling envelope, typically detected in the
 sub-millimeter and far-IR; Class\,I objects are evolved protostars
 surrounded  by an envelope and a circumstellar
 disk; Class\,II stars are
 objects with a circumstellar disk but lacking significant
evidence for a dense envelope and finally
 Class\,III stars
 are  objects that have lost their initial envelope and disk and show
 optical and near-IR colors
 consistent with normal photospheres.  
 By providing very sensitive 3.6-70\,$\mu$m data with good angular
resolution, {\it Spitzer} observations have provided
a much better census of the stellar populations in a number
of star-forming regions.  However, {\it Spitzer} data have also
complicated the YSO classification system, allowing detection
of some Class 0 objects at wavelengths as blue as 3.6\,$\mu$m.

 Our intent is  to compare the
 X-ray properties of protostars - where to date there are relatively
 few studies - with the X-ray properties of the more evolved
 Class\,II and III stars, for which the
 X-ray emission is well studied.  Such a comparison is fundamental in order 
 to understand the onset of 
 X-ray emission in YSOs and its evolution in the early phases of
 star formation.
 
Previous studies of X-ray emission from Class 0-I objects include \citet{ozaw05},
 who studied core F of the $\rho$\,Ophiuchi cloud and found
a high X-ray detection rate (64\%) of Class\,I sources. They  suggest that an
evolutionary effect in temperature and extinction 
 is present from Class\,I to Class\,III stars, these latter
having lower coronal temperatures and extinction. 

A similar result has also been found by \citet{giar07}. 
\citet{getm07} studied the IC\,1396N cometary globule where they found 6 X-ray
emitting protostars, including the Class\,0-I IRAS 21391+5802 which they claim to
be one of the youngest objects ever detected in X-rays. 
The comparison of X-ray emission from different evolutionary classes is, 
however, hampered by the difficulty of detecting X-rays from Class\,0 protostars.
For example, no X-ray emission has been found from Class\,0 stars
in the Serpens, one of the most sensitive surveys \citep{giar07}, even if
important detections have been found in other surveys
\citep[e.g.][]{tsub01,hama05,tsuj05,getm07}.


The ONC is a prime target for this kind of investigation. It is the nearest
concentrated site of star formation with a very large number of known members
covering all evolutionary stages.
It is the best studied region at all wavelengths, in particular, in the X-ray
band.
 Many interesting results have been obtained with the extraordinarily long 
{\it Chandra} observation performed by the COUP project. 
 To understand the evolution of X-ray activity from 
the very first stages of the star formation process,  it is necessary to
distinguish Class\,0-I objects from Class\,II stars. Such a classification,
based on the shape of the spectral energy distribution (SED) is not
available in the literature and 
requires complete and deep near and mid IR data.

We present here the results of the X-ray study of YSOs
 in the ONC classified using the {\it Spitzer} IRAC catalog \citep{mege05},
  groundbased JHK data (Robberto et al. 2007, in preparation)
 and the deep photometric catalog obtained within 
the HST Treasury Program \citep{robb05b}. 
Using the COUP data, we have studied the X-ray properties of YSOs
from their initial protostellar phase
(Class\,0-I objects) to the more evolved pre-main sequence (PMS) stage 
of no longer accreting and disk free stars (Class\,III objects). 
We have classified the ONC members using conservative criteria in 
order to minimize the contamination of the resulting samples.
Because of the large area coverage and the richness of the ONC, 
we have obtained an homogeneous and statistically significant sample
 of X-ray detected Class\,0-I 
objects. 
We are thus able, for the first time, to compute the X-ray luminosity function 
(XLF) of Class\,0-I stars and to compare it with the XLF of 
Class\,II and III stars.

\section{OBSERVATIONS}
  We have assembled a deep  photometric catalog
   of the ONC  from the UV
to  8\,$\mu$m obtained using new optical/NIR observations taken within 
the HST Treasury Program \citep{robb05b}, using a variety of telescopes and
detectors, and {\it Spitzer} IRAC \citep{wern04,fazi04} imaging.
The  optical photometry was derived  
using images taken with the  Wide-Field Planetary Camera 2
(WFPC2) of HST  \citep{robb05b}
and with the  Wide Field Imager (WFI) at the ESO 2.2\,m
Telescope (Robberto et al. 2007, in preparation).
 $JHK$ photometry was obtained from images taken    
with the Infrared Side Port Imager (ISPI) camera at the CTIO Blanco\,4m
telescope (Robberto et al. 2007, in preparation).
{\it Spitzer} IRAC photometry in the four standard filters centered on 3.6, 4.5, 5.8 and 
8.0\,$\mu$m was obtained by \citet{mege05}.
A detailed description of the optical-IR observations is 
given in Table\,\ref{observations}. We note that the magnitude limits given in
this table are typical values for each survey. However,
due to the complex structure of the nebulosity in the ONC (and due to
varying integration times for different parts of the region), the 
sensitivity is not uniform. 
For example the limiting magnitude at 3.6\,$\mu$m is about 16 within
most of the COUP
FOV, while it is about 10 within a region of 1\min\ in radius around the
brightest stars of the Trapezium cluster.

{\rotate
\tabcolsep 0.15truecm
\begin{table*}
\caption{Log of the new optical-IR observations used in this work.}
\label{observations}
\centering
\begin{tabular}{cccccccc}
\hline\hline
Nights&Telescope&Camera&Filters&Exp. time\tablenotemark{a}&Mag. Lim.\tablenotemark{b}& Field of & Reference\\
      & 	&      &       &	 [sec]         &	 &  view    & \\
\hline
2006 May      & HST	   &WFPC2& F336W\tablenotemark{c}&2$\times$400  &F336W$\simeq$20.0&30\min$\times$ 30\min& 1\\ 
2006 May      & HST	   &WFPC2& F439W\tablenotemark{c}&80  &F439$\simeq$20.7 &30\min$\times$ 30\min& 1\\ 
2006 May      & HST	   &WFPC2& F656N\tablenotemark{c}&400  &F656N$\simeq$18.0&30\min$\times$ 30\min& 1\\ 
2006 May      & HST	   &WFPC2& F814W\tablenotemark{c}&10  &F814$\simeq$20.0 &30\min$\times$ 30\min& 1\\ 
2005 Jan. 1-2 &ESO\,2.2\,m&  WFI   &U, B, V, I$_c$&3, 30, 280&I$_c\simeq$21.7&35\min$\times$ 34\min&2   \\
2005 Jan. 1   &CTIO\,Blanco\,4m&ISPI&J, H, K$_s$  &3, 30  &K$_s\simeq$18.7&6$\times$(10\Min25$\times$10\Min25)&2 \\
2005 Jan. 2   &CTIO\,Blanco\,4m&ISPI&J, H, K$_s$  &3, 30  &K$_s\simeq$18.7&8$\times$(10\Min25$\times$10\Min25)&2 \\
  & {\it Spitzer}\tablenotemark{d}       &IRAC & (3.6, 4.5, 5.8, 8.0)\,$\mu$m& &[3.6]$\simeq$16.8&5.6 sq. degrees&3  \\
\hline
\end{tabular}
\renewcommand{\footnoterule}{}  
\tablenotetext{a}{Exposure time per filter}
\tablenotetext{b}{Magnitude limit computed for stars with photometric errors smaller than 0.1}
\tablenotetext{c}{WFPC2 filters F336W, F439W, F656N, F814W corresponds to the standard filters U, B, H$\alpha$
and I, respectively}
\tablenotetext{d} {Data part of the {\it Spitzer}/GTO Programs 43 and 50 
taken in high dynamic range mode using 0.6 and 12 sec at each position.}
\tablerefs{(1) \citet{robb05b}; (2) Robberto et al. (2007) in preparation; (3) \citet{mege05}}
\end{table*}
}
\setlength{\voffset}{0mm}

\begin{figure} 
\centerline{\includegraphics[width=8cm]{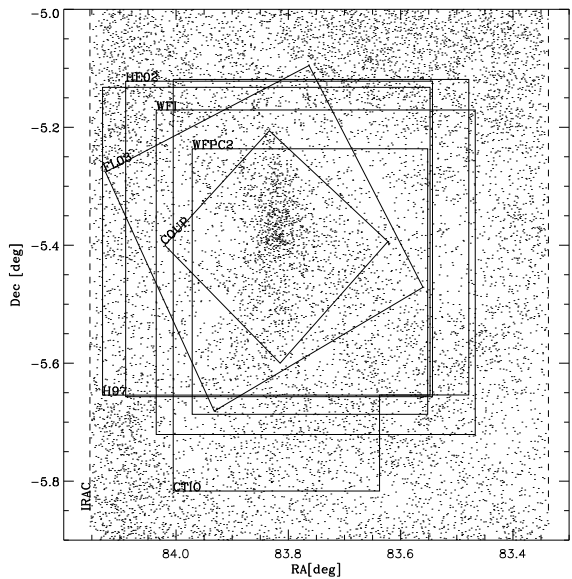}}
\caption{The  field of view (FOV) of the catalogs considered in this work
 are overplotted  on the
 spatial distribution of the
stars detected with IRAC. Note that the FOV 
of the IRAC (dashed line), \citet{rebu00} and 2MASS data 
(not indicated here) are larger than the field shown in this figure.}
\label{f01}
\end{figure}

 \subsection{X-ray data}
X-ray data used in this work were obtained with
the $\sim$838\,ks COUP observation of the ONC, taken with the ACIS camera that
covers a FOV of 17\min $\times$17\min. Details of data reduction and the
derived catalog of 1616 sources can be found
in \citet{getm05a}.
\subsection{Optical data}
New optical photometric data taken with the WFPC2 camera of HST
were used  by \citet{robb05b} to derive the optical photometric catalog 
used in this work. This catalog contains 1754 
selected point sources with at least a
detection in the F814 band.

The recent HST survey has been complemented with UBVI images taken 
with the WFI camera of the ESO\,2.2\,m Telescope. Details on data reduction and
photometric catalog derivation will be described in Robberto et al. (2007, in
preparation). This catalog contains 2744 selected point sources with at least a
detection in the I band.

Additional optical data were retrieved from literature including the catalogs of 
\citet[][ hereafter H97]{hill97}, \citet{rebu00} and \citet{herb02}.
\subsection{$JHK$ data}
The $JHK$ photometric catalog has been obtained using images
 taken with the ISPI camera at the CTIO Blanco\,4\,m
Telescope. Details on data reduction and photometry will be given in 
Robberto et al. (2007, in preparation). 
The survey includes a mosaic of 11  fields, each of
10\Min25$\times$10\Min25, centered 
around the Trapezium cluster.
All fields were observed once using a 5 point dithering pattern. 
From the source catalog
we excluded  saturated objects, that is, those with a peak flux in each band
 higher than 37\,000 counts and entries with errors in magnitudes larger than 
  0.2. After this selection, we 
 were left with 17\,824 entries in the table.

Finally, we constructed a consolidated catalog of individual objects by 
merging entries with positions within 0\Sec6, the radius at which
spurious identifications start to dominate.
For each object, we computed the mean coordinates 
and magnitudes. 
 Our final $JHK$ photometric catalog contains  13\,987 objects.

Additional $JHK$ data were retrieved from the  2MASS catalog
\citep{cutr03}. 

\subsection{{\it Spitzer} IRAC data}
A description of the data, the reduction procedure and the derived 
photometric catalog obtained from the {\it Spitzer} IRAC images is given in
\citet{mege05}.  The survey covers the Orion A and B molecular clouds for a
total of 5.6 sq. degrees with a sensitivity of 17.0, 16.0, 14.5 and 13.5 mag in
the 3.6\,$\mu$m, 4.5\,$\mu$m, 5.8\,$\mu$m and 8.0\,$\mu$m  bands, respectively.
The uncertainties in the calibration are about 5\%, but for a few stars are
significantly worse due to the bright nebulosity in 
 the 5.8 and 8.0\,$\mu$m bands. 
 Within the region  studied in this work, i.e. the COUP FOV,
the catalog includes 
1244 sources 
 detected both at 
3.6\,$\mu$m, 4.5\,$\mu$m and 334 
sources detected both at 
 5.8\,$\mu$m, 8.0\,$\mu$m. 
 The {\it Spitzer} Orion Survey also included  MIPS
 observations at 24\,$\mu$m that we do not use in the following analysis 
 since they are saturated in the COUP region due to the high
 background from the HII region.

\subsection{Cross identifications of sources}
All the photometric catalogs obtained from the observations
described in Table\,\ref{observations} were assembled in a single
database.
Common objects between two catalogs
were found  using the nearest neighbour method by first considering
two sources as the same object if their positions agree to within
a specified amount -
1\sec-3\sec -  depending on the
absolute astrometric accuracy of the catalogs.
Matched sources  were then used to compute the median and the
standard deviation of the offsets.
These median values were used to remove
the systematic offsets between each
pair of matched catalogs,
and new matches were performed, for all pairs of catalogs, using new tolerances  
computed using the standard deviation of the offsets. 

Using the same matching procedure we have cross-correlated 
these data with the  
literature catalogs of \citet{hill97},
 \citet{rebu00}, 
 2MASS \citep{cutr03}, \citet{herb02}, \citet{muen02} and
 the X-ray source list from the COUP Survey \citep{getm05a}.

Finally we collected
all data from all the cross-matched catalogs, including all objects detected in
at least one catalog.
  
The  field of view (FOV) covered by these  surveys
are overplotted in Fig.\,\ref{f01} on the
 spatial distribution of the
stars detected with IRAC around the ONC. 
Note that the FOV of the IRAC, \citet{rebu00} and
2MASS data (not indicated)
 are larger than the field shown in the figure.

Since this work is focused on the X-ray properties of
the young stars in the ONC, only stars within the COUP FOV will be considered 
in the following  analysis. In addition, 
photometric data   with errors smaller 0.2 mag will be considered.



\section{YSO CLASSIFICATION \label{classification}}
In this Section we describe how we select ONC stars belonging to the three
different evolutionary classes 0-I, II and III. Our approach is to select
conservative, even if incomplete, samples in order to be able to
compare the X-ray properties
of the stars in different evolutionary stages using samples
that are as ``uncontaminated" as possible.
 

\subsection{Class\,0-I protostars}
According to \citet{lada84}  and \citet{adam87}, the first stage of YSO 
evolution is the infall  phase around a central protostar undergoing
accretion, characterized by
strong emission in the sub-millimeter and far-infrared and weak emission
shortward of 24\,$\mu$m. These objects, indicated as Class\,0, 
can have outflows and are defined as 
 protostars with
half or more of their mass in the envelope. More evolved protostars are the
Class\,I objects
with both a disk and envelope surrounding the central star.
Both these classes of protostars are characterized by
 a steeply rising SED in the infrared range,
 although pure Class\,0 objects are expected to be   
  fainter at near and mid-IR bands \citep{alle07}.

Since our data extend only to 8\,$\mu$m,
 we cannot distinguish between Class\,0 and Class\,I stars and we will
 combine them into one group as Class\,0-I protostars;
we classify  them using the following criterion:
we computed ${\rm log}(\lambda F_\lambda)$ 
over the wavelengths
$[\lambda_1,...\lambda_5]=[2.2,3.6,4.5,5.8,8.0]\,\mu$m,
 i.e. the K-band and the four IRAC bands with
the \citet{lada06} flux conversion factors that are in agreement with 
the \citet{reac05} values; 
for each star we then computed the
slopes for each pair of adjacent bands
$$\frac{{\rm log}(\lambda_{i+1} F_{\lambda_{i+1}})-
	{\rm log}(\lambda_{i}   F_{\lambda_{i}})}
	{{\rm log}\lambda_{i+1}-{\rm log}\lambda_{i}}~~~ {\rm for}~~i=1,4;$$
 we consider Class\,0-I objects those for which the slopes are all
larger than 0, i.e. objects with rising SEDs. This 
 criterion
is more conservative
than using the single slope (usually indicated as spectral index)
computed from the linear fit of the 
${\rm log}(\lambda F_\lambda)$ function over the whole wavelength range,
because it only selects monotonically increasing 
SEDs.
Moreover,  it is more robust
than those based on only a
color-color diagram (CCD)  since it is based
simultaneously on all the
 known magnitudes. 

Our sample of Class\,0-I objects includes all the objects for which we have at least
three of the five magnitudes and therefore at least two
slopes. This allows us to include objects detected only in the K, 3.6\,$\mu$m 
and 4.5\,$\mu$m
bands but not at 5.8 \& 8.0\,$\mu$m; because of the bright nebular background 
and the lower sensitivity 
 at 8\,$\mu$m,
 many objects
are not detected in these two bands (roughly 44\% 
of the Class 0-I sample).

\begin{figure*} 
\centerline{\includegraphics[width=18cm]{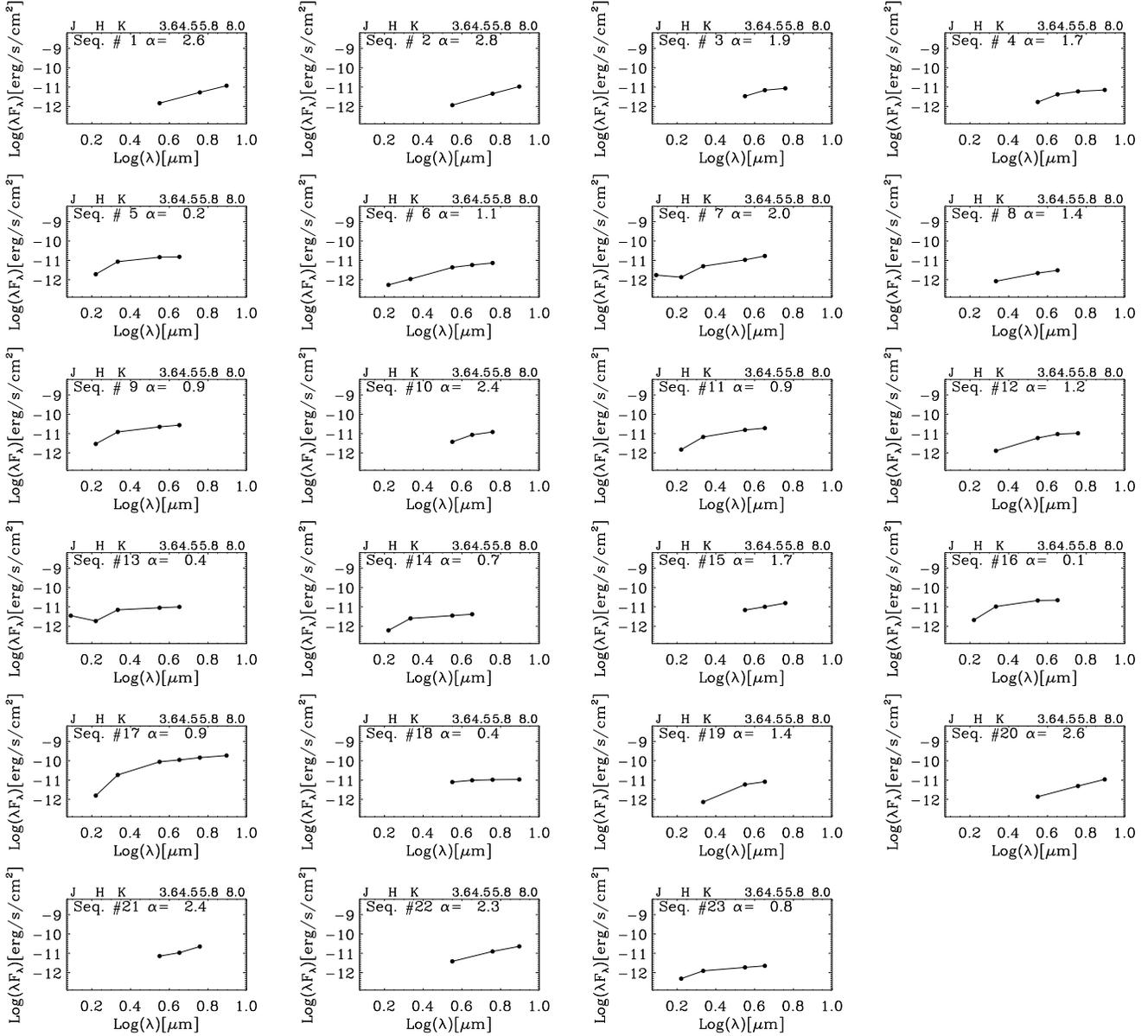}}
\caption{SED of the 23 Class\,0-Ia candidate objects with a rising SED from K to
8.0\,$\mu$m. The sequential number in our catalog and the spectral index
computed using the available IRAC magnitudes  are also indicated.}
\label{f02}
\end{figure*}

 Using this criterion, we 
 end up with 41  potential Class 0-I objects in the COUP field of view
 \citep{getm05a}.
  Eighteen
 of them have a detection
  in at least one of the optical 
 UBVI bands. Since protostars are expected
  to be heavily embedded and therefore highly reddened objects,
  we do not expect to detect them 
   in the visible.
  Of these 18  peculiar objects,
5 (COUP \# 693,     747,     826,     948,    1011) 
are in the list of the X-ray detected proplyds given in
\citet{kast05}. From a visual inspection of ACS/HST images, we conclude that
all 18 of these objects have  a counterpart in the HST optical
bandpasses, while the other 23 objects are not detected at these bands.
We eliminate these 18 objects from our list of candidate Class\,0-I stars
 and we consider  
the remaining 23 stars  as Class\,0-Ia objects to distinguish them from the
additional sample  of candidate Class\,0-I described below.
 The SEDs of the 23 selected Class\,0-Ia stars are plotted  in 
Fig.\,\ref{f02}.  The photometry is given in 
Table\,\ref{cl0I_1} where Column\,1 gives the sequential number in our catalog,
Column\,2 is the COUP identification number, Columns\,3 and 4 are the J2000
coordinates, Columns\,5 to 11 give the JHK and IRAC magnitudes.
 Of these 23 objects, 
10 ($\sim$40\%) have an X-ray counterparts
 in the COUP data. 
One of these (COUP \# 702) is in the \citet{getm05b} 
 list of nonflaring 
 COUP X-ray sources without optical and NIR counterpart  and was  
  flagged by them as a
probable extragalactic object (EG).

\begin{figure*} 
\centerline{\includegraphics[width=19cm]{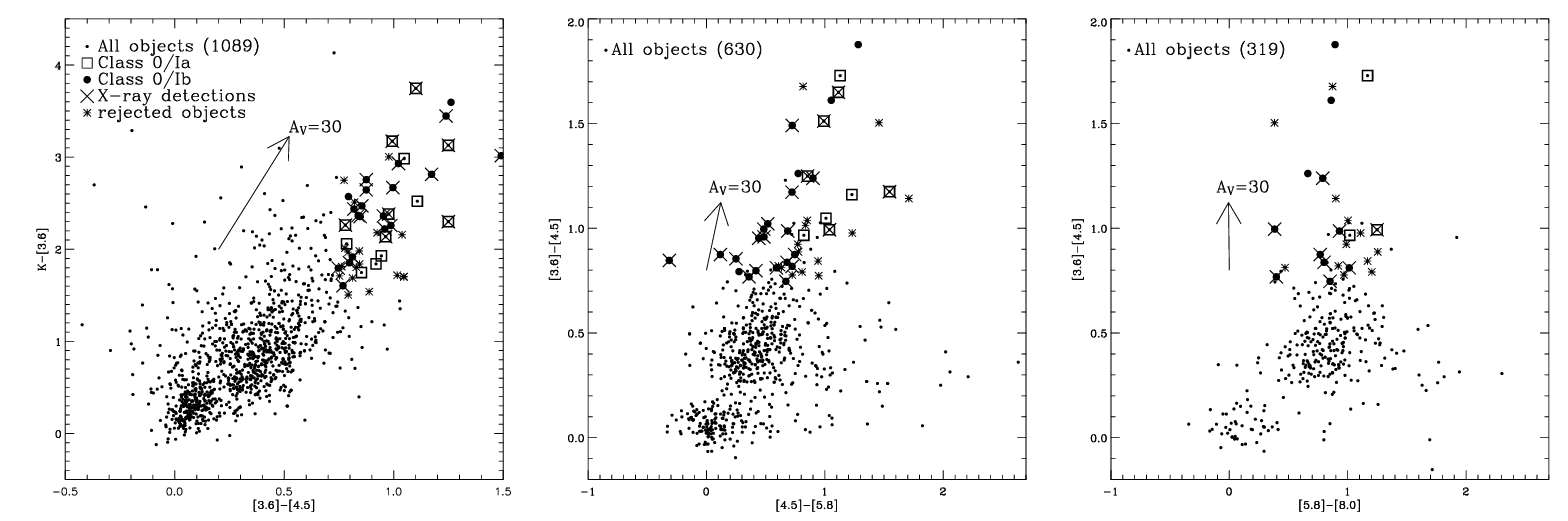}}
\caption{CCD of the stars in the COUP field of view (dots)
created using the $K$ and the
{\it Spitzer} IRAC magnitudes. 
The total number of objects in each diagram is reported in the three panels.
Squares and  large filled circles represent 
the   Class\,0-Ia and 0-Ib objects, respectively, 
 while
 {\tt X} symbols indicate
  those with an X-ray counterpart; asterisks
indicate objects rejected because they have an optical counterpart.
The reddening vectors computed using the IR reddening law 
given in \citet{flah07} for Orion A are also plotted.}
\label{f03}
\end{figure*}

Fig.\,\ref{f03} shows the color-color diagrams
(CCDs) obtained using the K and IRAC magnitudes 
of all the stars in the COUP field of view.
 The plot on the left panel includes more
stars than those  on the central and right ones 
because it does not require detections at 5.8 and 8.0\,$\mu$m. 
The limited sensitivity of the 8.0\,$\mu$m channel means that few
objects with zero color (consistent with photospheric emission) appear
in the right panel. 
The selected  Class\,0-Ia objects with increasing SED,
indicated in Fig.\,\ref{f03} by squares, show [3.6]-[4.5]$>$0.7
and [5.8]-[8.0]$\gtrsim$1.1. The objects with [3.6]-[4.5]$>$ 0.7
and [5.8]-[8.0]$\lesssim$1.1 that are usually also classified as Class\,0-I 
stars \citep{alle04,mege04}, show a decreasing or flat SED at wavelengths longer
than 4.5\,$\mu$m. 
The spectral energy distributions of protostars in this
spectral regime are complex as
rotation \citep{keny93} and outflow cavities
\citep{whit03} reduce the
extinction toward the poles of the protostellar envelopes allowing
radiation to leak  and scatter
out.  Consequently, protostars
can also exhibit decreasing slopes, particularly between 4.5 and 8.0~$
\mu$m, depending on their
inclination \citep{whit03}.  Furthermore, highly
reddened Class II
objects can have increasing slopes; however, these slopes will
generally flatten or decline for
$\lambda \ge 4.5 \mu$m due to the flattening of the reddening law
and the broad silicate feature
at 9.7~$\mu$m \citep{flah07}.  Thus, although the IRAC data
can enhance the identification
of protostars by extending the SEDs to 8 microns; establishing a
sample of protostars uncontaminated by
reddened Class II objects is still difficult  without photometry at
wavelengths $\lambda > 8$~$\mu$m.

\begin{figure*} 
\centerline{\includegraphics[width=18cm]{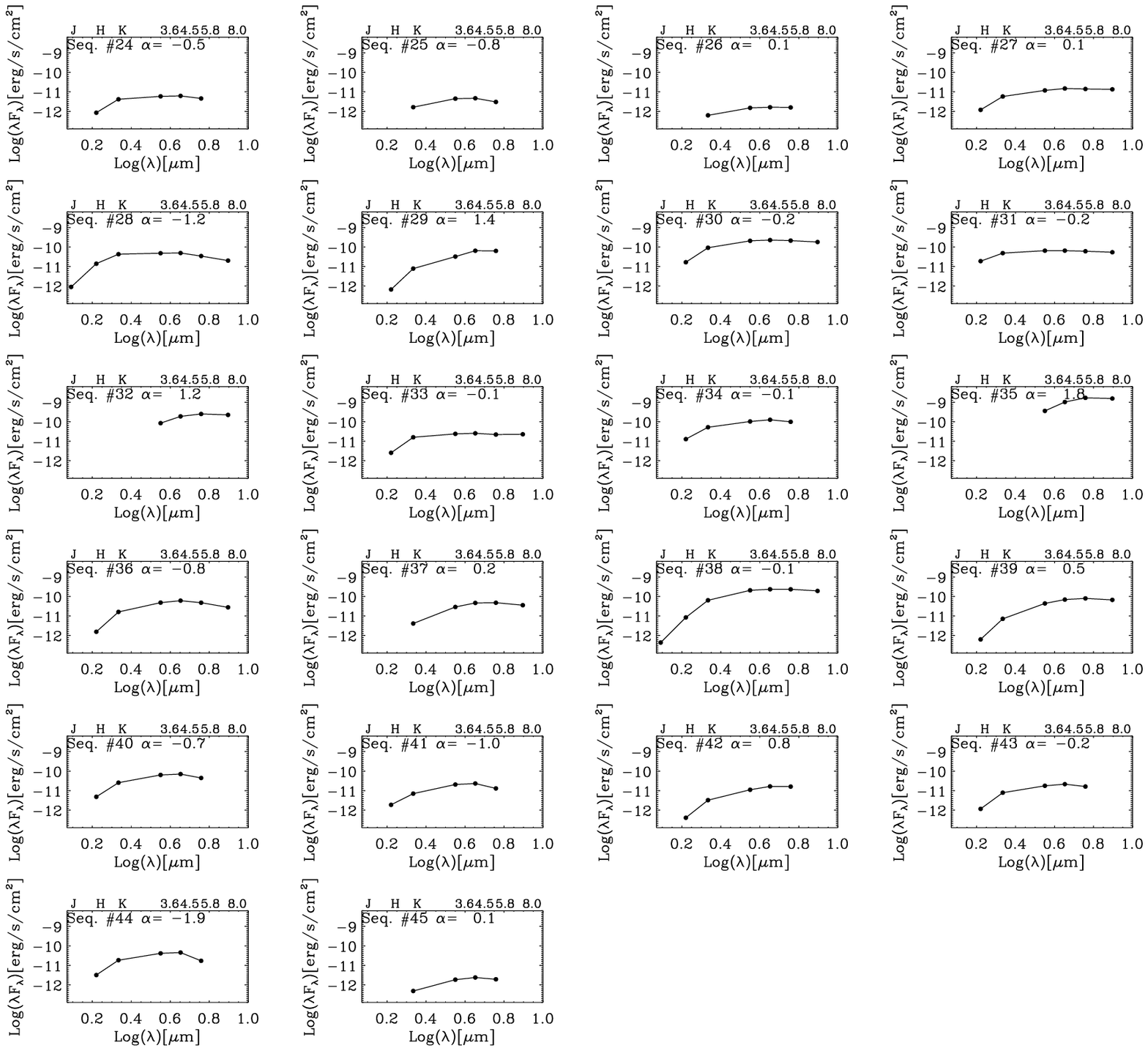}}
\caption{SED of the 22 Class\,0-Ib candidate objects with a rising SED
from K to 4.5\,$\mu$m. The sequential number in our catalog and the 
spectral index
computed using the available IRAC magnitudes  are also indicated.}
\label{f04}
\end{figure*}

To take into account these cases, we also include the objects with  a 
decreasing SED for $\lambda > 4.5 \mu$m
in our Class\,0-I sample.
There are 28  such objects in total. 
After rejecting the six objects with optical counterparts in the HST imaging, we 
are left with 
22 objects,  which we
refer to hereafter as Class\,0-Ib sources. 
 The  SEDs of the 
 Class\,0-Ib objects are shown in
 Fig.\,\ref{f04}, while the photometry is given in 
Table\,\ref{cl0I_2}, 
analog
 to Table\,\ref{cl0I_1}.
 Among these objects,   18 
($\sim$80\%) are in the COUP list of
 X-ray detection, including  COUP source 1565, which was classified as an 
 extragalactic object by \citet{getm05b}. 
 
If we consider both 
Class\,0-Ia and Class\,0-Ib samples,
our list of X-ray detected Class\,0-I objects (10+18)
  is one of the largest  sample of Class\,0-I stars with 
 evidence of X-ray emission \citep{getm07,giar07}.
 
Our selected  Class\,0-I
objects fall in the region
of the envelope models indicated
in \citet{alle04} and \citet{mege04}. 
There are
a small number 
 of objects falling in the Class\,0-I 
regions of the CCD which were
not selected with our criterion because either
they have an optical counterpart (18+6 in total) or 
their SED band-to-band slopes do not increase as required by our
criterion.
Whereas some of these rejected 
objects could be Class\,0-I stars, by adopting a conservative selection 
criterion, we 
have minimized
the adverse effect of sample contamination 
 for our comparison of X-ray properties.

A careful examination of the near-IR and IRAC
colors of our Class\,0-I candidates (see Figs.\,\ref{f03}
 and \ref{f05})
suggests to us that the Class\,0-Ia's (on average) are
inconsistent with their being heavily reddened Class II's.
In particular, the bottom-right panel of Figure 5 shows
that, using the Orion A reddening law of Flaherty, the
Class\,0-Ia's are too red in [4.5]-[5.8] to be heavily
reddened Class\,II's - and in any case would require
reddenings of A$_{\rm V}\geq60$.  The bottom left panel of
Fig.\,\ref{f05} also indicates that the Class\,0-Ia's, on average,
would need very large reddenings (A$_{\rm V}\sim50$) for them to
be Class\,II sources.  Such heavily embedded objects would
likely be very young in any case - and we therefore believe
it is more likely that they are, indeed, Class\,0 or Class\,I
objects.   By contrast, the Class\,0-Ib objects are located
in all of the diagrams in Figs.\,\ref{f03}
 and \ref{f05}  in positions
consistent with being moderately reddened (A$_{\rm V}\sim30$
on average) Class\,II's.   We suspect the Class\,0-Ib's
are a mixture of true Class0 or I objects, and heavily
reddened Class\,II's - without more data we cannot
clean the sample any further than we have.

 Another possible worry is that our Class\,0-I candidate sample includes 
 systems seen edge-on in or around the HII region. In these cases, 
 relatively evolved Class\,II objects may display a SED peaking at 
 mid-IR and far-IR wavelengths, as suggested by \citet{robb02}.
  However, since edge-on
 disks should be randomly distributed in the nebula, 
 clustering in the dark regions
 of the selected objects (see Fig.\,\ref{f06}a) strongly suggests
  that they are
 real protostars.
 

\subsection{Class\,II stars \label{classII_sec}}
Following \citet{hart05}, Class\,II objects within the COUP field of view
were selected using the condition $0.2<[3.6]-[4.5]<0.7$ and
 $0.6<[5.8]-[8.0]<1.1$. These criteria are slightly more restrictive than 
 those adopted by \citet{mege04} in order to attempt to create
 as "pure" a set of Class II's as possible.
 With our selection criteria,  
 we may lose reddened Class\,II
 objects, but assuming that X-ray properties of Class\,II
  stars are independent of their reddening, our sample should be representative
  of the Class\,II population. 
     
  Using this criterion, we find 148 
 Class\,II stars with NIR photometry
given in Table\,\ref{clII},
 analogous to Table\,\ref{cl0I_1}; 130 of these objects (about 
88$\%$) are X-ray  detected. One of them, the COUP source 1401, is listed by
\citet{getm05b} as a  flaring source and  
classified as an Orion Molecular Core (OMC) member, 
whereas another Class\,II object,
 COUP  source 
 751, was classified by the same work as  extragalactic object. 
 Among the 130  
X-ray detected Class\,II stars
with a counterpart in the \citet{hill97} catalog,
83  have a measure of the equivalent width of the 
Ca\,{\small II} 8542\AA\ line $W$(Ca\,{\small II}). Among these,
28 have $W$(Ca\,{\small II})$<-$1, 
which is the limit used by \citet{flac03a} to classify strong accretors.
Therefore we have about 55 
objects (i.e. $\sim$66\%) 
that have a near-IR bright inner disk but relatively low 
accretion rate.   
Of the 148 Class\,II stars we have identified, we will consider 
 in the following analysis only 
the 87 X-ray detected  and the 9 X-ray undetected stars for which 
  masses are known \citep{hill97,getm05a}.
We retain only these
stars  in order to study the
X-ray properties as a function of stellar mass.

\subsection{Class\,III stars \label{classIII_sec}}

 Class\,III objects are PMS stars that do not show IR
excesses and therefore 
were selected as those having IRAC colors near zero.
In order to include all the stars not detected in the longer
 wavelength IRAC
bands, we selected all the stars with $K-[3.6]<0.5$ and
 rejected those with $[3.6]-[4.5]>0.2$ or $[4.5]-[5.8]>0.2$ or
$[5.8]-[8.0]>0.2$, when these latter colors were available,
 i.e. objects with IR excesses. Stars 
 undetected in the IRAC bands at wavelengths
longer than  3.6\,$\mu$m were included since 
their non-detection in these  bands
is compatible with purely photospheric SEDs.

Using these criteria, we selected 205 objects,
 with photometry 
given in Table\,\ref{clIII}; 160 of  them
 (about 80\%)
have an X-ray counterpart.
 Among these latter, 150
are in the list   of
\citet{hill97}. In particular, 130 of them have
 a proper motion membership
probability larger than 90\%, 6 have a membership probability equal to 0 and 
the other 14 have no  membership probabilities. 
For the remaining 10 stars without
a counterpart in the list  of
 \citet{hill97}, 
 X-ray detection is the only known 
membership criterion. 
We note that none of these 10 objects is either in the list of flaring
or  in that of non-flaring sources of \citet{getm05a}.
 Therefore
 the membership class for these objects 
 remains uncertain.
 
 Of the 45 objects without an X-ray detection, 18 have
a counterpart in the \citet{hill97} catalog but only 5 of these have a proper
motion membership probability larger than 90\%. 

The remaining 
40 objects without a membership criterion
could be  either Class\,III stars without X-ray activity, or
 field stars. 
Of the 205 Class\,III stars initially selected, we 
will consider the 103 
X-ray detected stars for which the mass is known 
\citep{getm05a} and with \citet{hill97} membership probability larger than 90\%;
in addition, from the sample of X-ray
 undetected Class\,III candidates, we will retain
only the 2 objects with membership probability larger than 90\%
and for which  a mass has been derived by \citet{hill97}.
 
\begin{figure*} 
\centerline{\includegraphics[width=15cm]{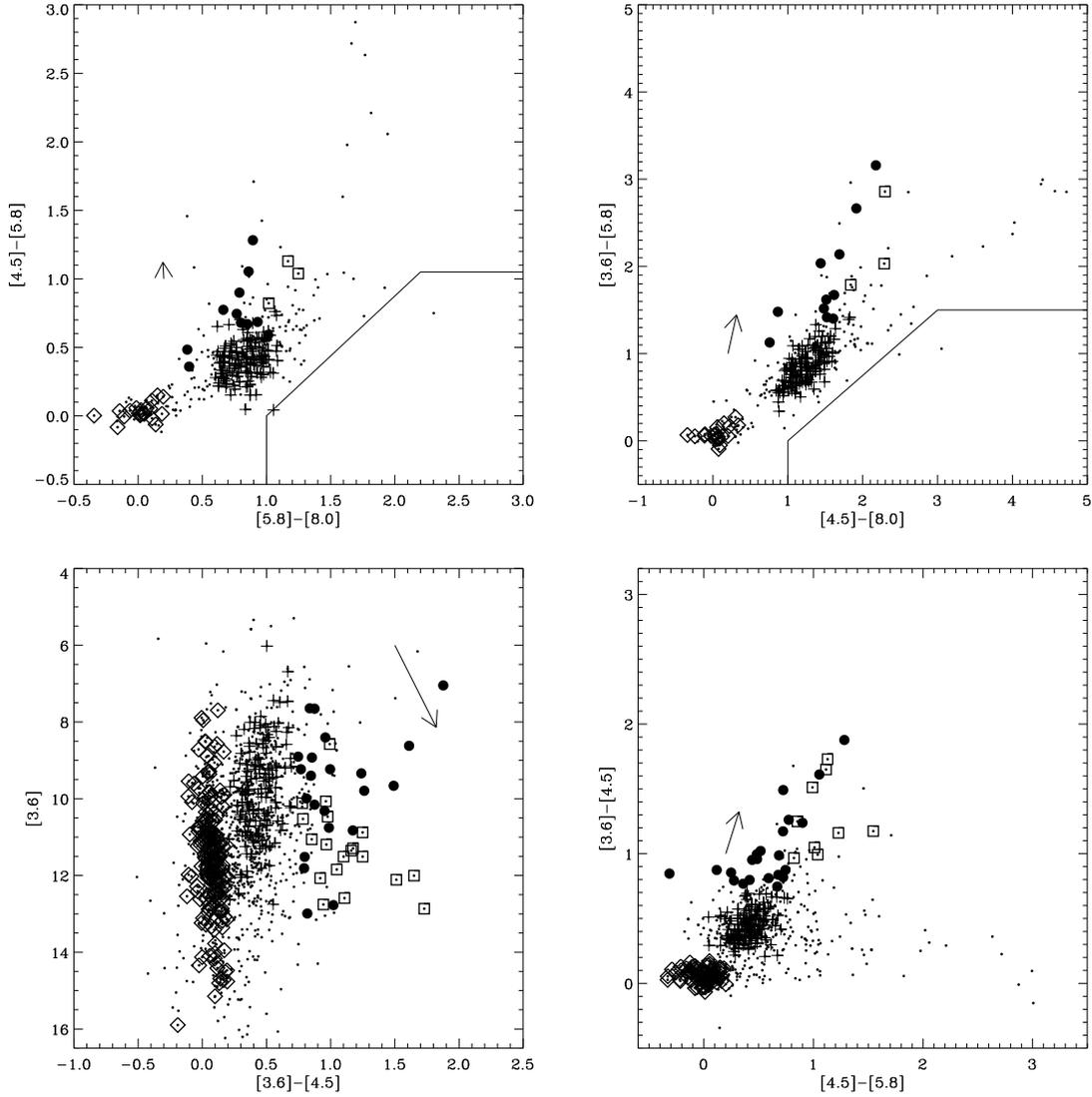}}
\caption{IRAC CCD and color-magnitude diagrams of the  different
populations in the ONC.  Dots indicate
all the stars in the COUP field of view with detections at the bands used in
each panel. Squares  and large filled circles
are the Class\,0-Ia and 0-Ib objects, respectively,  plus symbols 
 are the stars classified  as Class\,II and  diamonds are
the Class\,III stars.  The reddening vector corresponding to A$_{\rm V}=30$,
computed using the \citet{flah07} reddening law, is plotted on each panel.
 The solid lines 
in the  above panels separate the region dominated by YSO's
from the region expected to be dominated by galaxies.
}
\label{f05}
\end{figure*}

Fig.\,\ref{f05} shows the selected stars
in the IRAC CCD and color-magnitude diagrams. 
These diagrams show that our samples of YSOs most likely do not include AGN or 
star-forming galaxies;
that is, they do not include objects falling in the wedge-shaped
regions outlined in the upper panels of Fig.\,\ref{f05}.
Star-forming galaxies are usually dominated by PAH  features in
the 5.8 and 8.0\,$\mu$m  and therefore
have $[5.8]-[8.0]>1.0$ but $[3.6]-[4.5]\lesssim0.3$, that corresponds to 
the regions indicated
by solid lines 
in the two upper panels of Fig.\,\ref{f05} \citep{gute07}.
AGN, instead, have 
$[3.6]-[4.5]$ and $[5.8]-[8.0]$ colors very similar to the Class\,II YSOs,
but are typically much fainter than YSOs, having, for example,
[4.5] magnitudes  \citep{mege05},
 much fainter than those of
the selected YSOs (see  bottom-left panel in
Fig.\,\ref{f05}).

\section{SPATIAL DISTRIBUTION}
\begin{figure*} 
\centerline{\includegraphics[width=18cm]{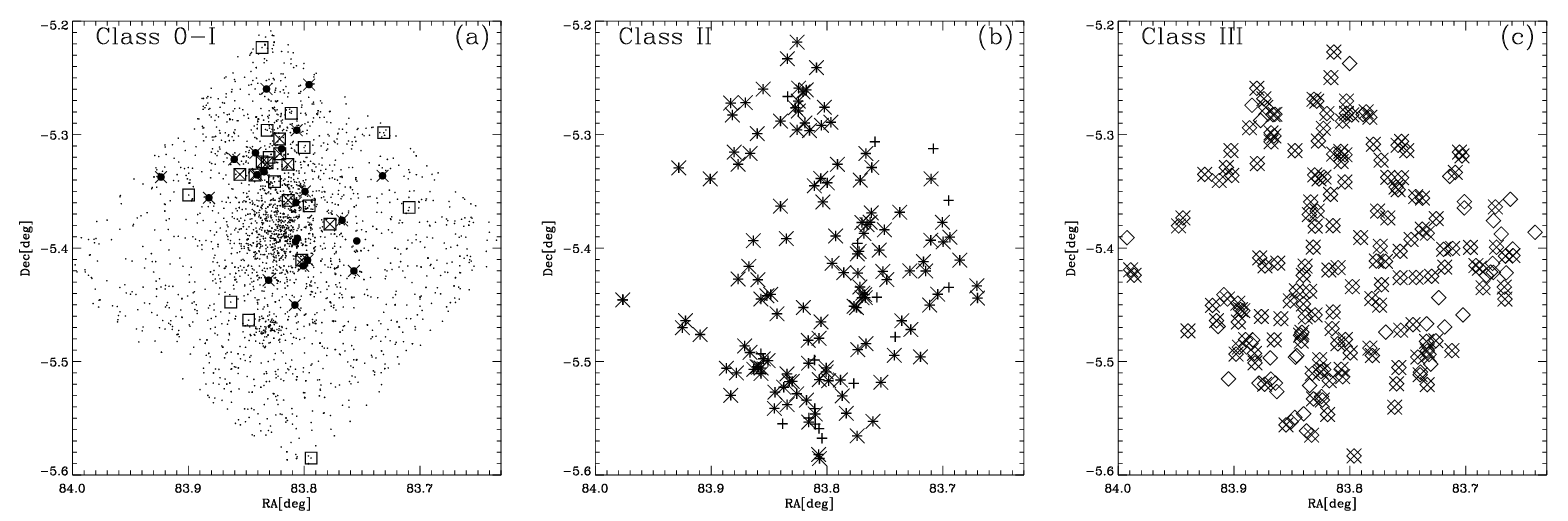}}
\caption{Spatial distribution of all the objects in
our catalog within the COUP field of view that have been detected in at least
two of the UBVIJHK and IRAC bands. Squares  and  large filled 
circles indicate
Class\,0-Ia and 0-Ib objects,  plus symbols  
the   Class\,II stars 
and  diamonds  the   Class\,III objects. 
In all the panels, X-ray detected objects
are indicated with an {\tt X} symbol. }
\label{f06}
\end{figure*}

The spatial distribution of YSOs within the molecular cloud from which
they have formed
may give  important clues for understanding the star formation
process. Using the classification 
 defined in the previous section, we have analyzed the spatial distribution
of the YSOs in the ONC, as shown in 
Fig.\,\ref{f06}a where we plot   the 
position of all  2590  objects in
our catalog within the COUP field of view that have been detected in 
at least two of the UBVIJHK and IRAC bands and the Class\,0-I candidates.
 Stars of Class\,II and III  
are plotted in Fig.\,\ref{f06}b and c, respectively.

\begin{figure*}[t] 
\centerline{\includegraphics[width=18cm]{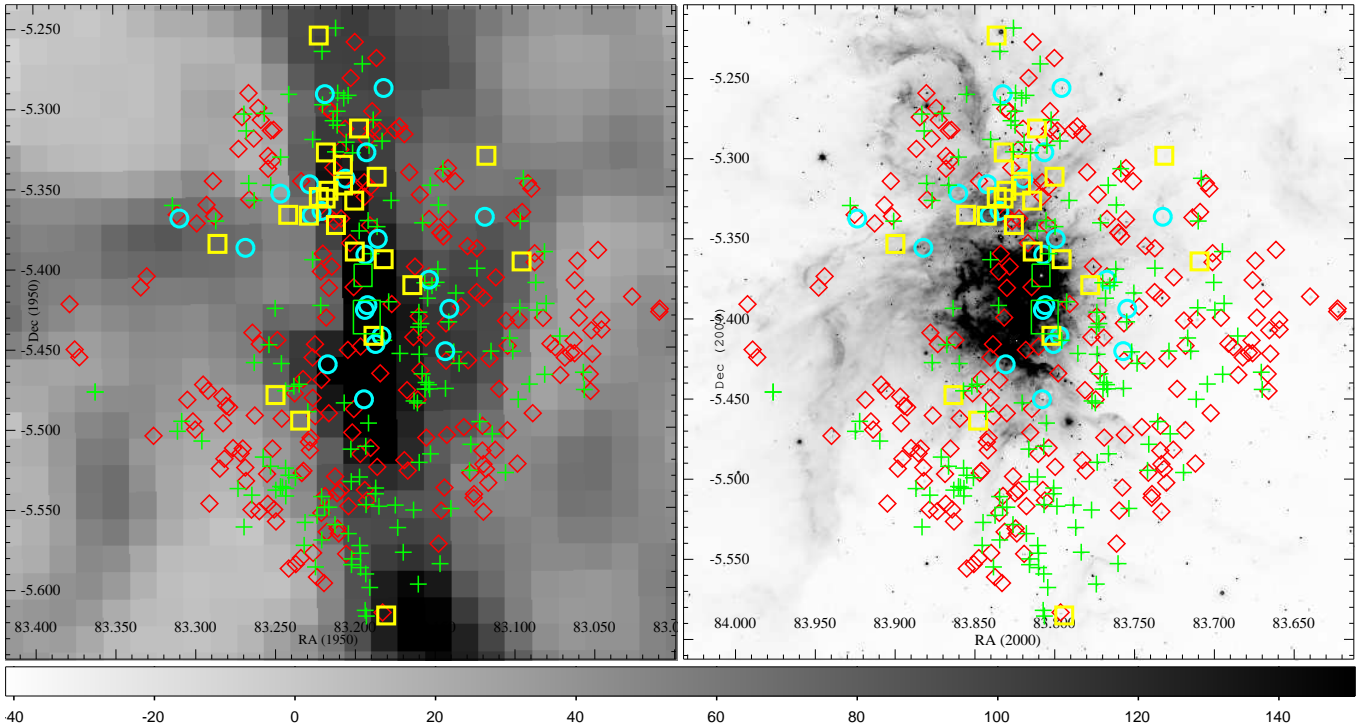}}
\caption{{\it (Left)} Velocity integrated $^{13}$CO emissivity from
\citet{ball87}  in the region around the Trapezium cluster and
{\it (right)} {\it Spitzer} IRAC image at 3.6\,$\mu$m in the same region. 
Both images are shown with inverted color map. Yellow squares and  light blue
 circles
 indicate Class\,0-Ia and 0-Ib objects,  green plus symbols
are Class\,II stars while red diamonds are Class\,III stars.
The 40\sec$\times$50\sec upper green box indicates  
the BN-KL  region while the 60\sec$\times$75\sec   bottom green box 
indicates  the Orion Molecular Cloud 1 South (OMC-1S).}
\label{f07}
\end{figure*}

Figure\,\ref{f07} shows the velocity integrated $^{13}$CO map obtained by
\citet{ball87}   and the {\it Spitzer} IRAC image at 3.6\,$\mu$m 
  where the
 different classes selected in this work have been overplotted.
The  upper green box indicates a 40\sec$\times$50\sec\ region around 
the Becklin-Neugebauer object  and the
Kleinmann nebula, also indicated as BN-KL 
(RA$_{2000}$=05$^h$35$^m$14\timesec16,
Dec$_{2000}$=-05\deg22\min21.5\sec). 
The bottom green box
 indicates a  60\sec$\times$75\sec\ region 
around the Orion Molecular Cloud 1 (OMC-1S) 
(RA$_{2000}$=05$^h$35$^m$14\timesec50,
Dec$_{2000}$=-05\deg25\min49.0\sec).
These boxes show the limits of the areas studied in \citet{gros05}. 

Most of the  Class\,0-I stars are  
   concentrated in the central and north-east parts
of the COUP field of view  
  along a region running north-south from the
Trapezium that 
follows the north-south distribution of the $^{13}$CO emission, i.e. of the 
dense star-forming gas. The distribution of our selected Class\,0-I objects
 is biased by our NIR observations which are more sensitive
towards the north of the molecular ridge where the nebular
contribution is lower. The lack of Class\,0-I objects in
the BN-KL and OMC-1s regions is due to the bright nebulosity, as 
is evident from the image at 3.6\,$\mu$m, that causes a decrease of sensitivity 
in the central region. For this reason, the spatial
 distribution of our Class\,0-I objects 
 cannot be directly compared with that obtained by 
 \citet{lada00,lada04}, although in fact the two spatial 
 distributions are quite similar.


Figures\,\ref{f06} and \ref{f07} show the same structure already
found in \citet{feig05} using  the lightly X-ray 
absorbed stars: they interpreted this distribution as due to a process of 
violent relaxation.  
Note that, the latter study examined the spatial distributions
stratified by absorption, while here the stars are stratified by individual YSO
evolutionary classification. The features of the spatial distribution
visible in Figs.\,\ref{f06} and \ref{f07} include: the OMC-1S core,
BN/KL core and northern extension which are part of the larger filament along
the middle of the Orion A cloud and finally the East-West asymmetry that in
\citet{feig05} was
attributed to non-equilibrium dynamics. The latter only shows up in Class\,III
systems.

 The spatial distributions of  
Class\,II and III stars are also 
in agreement with the spatial distribution found by 
\citet{hill97} who,  using \citet{dant94} isochrone ages, 
 found that youngest stars are clustered toward the cluster
center  while slightly older stars are more widely distributed. 
The  elongated asymmetric 
distribution of our candidate Class\,0-I stars and the large-scale EW
asymmetry are consistent with the
distribution derived by \citet{hill98} who concluded
 that the structure of the ONC is more the result 
of the different environmental conditions rather than the result of dynamical
evolution.  

 There is
a lower percentage of Class\,II objects in the central region,
but this could be due to the reliance on the 8.0\,$\mu$m detections for
 the selection of Class\,II stars.
Our spatial distribution are, therefore, strongly dependent on the 
sensitivity of our NIR data which is quite limited toward the central region,
where the contamination by the bright 
OB stars and nebular emission is very high.

%

\section{EVOLUTION OF X-RAY ACTIVITY}

The very deep X-ray COUP data
together with the relatively large sample of Class\,0-I objects selected with
the {\it Spitzer} observations 
 allow us, for the first time, to analyze and compare, 
in a statistically significant way, the X-ray properties of the 
YSOs in  their initial protostellar phase 
with those of more evolved PMS stars. 

As already mentioned in Section\,\ref{classification}, we find that
the fraction of Class\,0-Ia objects detected in the X-rays
is almost 40\% (10/23),  that of 
 Class\,0-Ib  and   Class\,III  
    is about 80\%  (18/22 and 160/205, respectively) while that of
Class\,II objects
is  about 88\%( 130/148).  The analysis described in this section aims at
understanding whether
 the X-ray emission can be related to the IR properties of YSOs, which we
 take as signposts for 
the different phases of protostellar and PMS evolution.   

\subsection{X-ray luminosities \label{xluminosity}}
We use the results of the spectral analysis presented by \citet{getm05a} to
compare the X-ray properties of the  different populations of YSOs in the
ONC.
For the X-ray luminosities of
detected stars we adopt  
the absorption corrected luminosities in
 the total 0.5--8.0\,keV band, L$_{t,c}$, from Table 8 of \citet{getm05a}.

In order to compare unbiased luminosity functions  of the  
different classes, we derive  upper limits to
the X-ray luminosities of the X-ray undetected objects.
Upper limits to the photon count rates of the 13+4 undetected Class\,0-Ia 
and Class\,0-Ib objects were calculated 
with PWDetect \citep{dami97b,dami97a} 
consistently with the source
 detection procedure of the other COUP sources \citep{getm05a}, adopting a
 detection threshold significance   of 5\,$\sigma$.
 The upper limits for the 9 undetected Class\,II stars and 
 for the 2 undetected Class\,III stars
  were taken from Table\,11 of \citet{getm05a}.
 

In order to convert upper limits to the count rates into unabsorbed 
 X-ray luminosities, we  derive
an extinction dependent conversion factor. 
The CCDs in Fig.\,\ref{f03} indicate that
 X-ray detected and undetected Class\,0-I objects have similar reddening.
 We therefore computed the conversion factors of the X-ray detected 
 Class\,0-I objects as the ratio between the unabsorbed X-ray luminosities 
 (L$_{t,c}$) and the count rates  taking into
 account the PSF fraction within the extracted area and the effective exposure
 time at the source position \citep{getm05a}, and we plotted these values as a
 function of the N$_{\rm H}$\footnote{The  H column densities were derived
  by \citet{getm05a}
 from the spectral analysis of the COUP sources.} values.
 We  assume for the X-ray undetected Class\,0-I objects a conversion
 factor computed as the
 median value of the conversion factors of the 
 X-ray detected Class\,0-I objects, which corresponds for these  stars to 
 $N_{\rm H}\simeq4.5\times10^{22}$cm$^{-2}$. 
 However,
 in order to check whether  our results are robust with respect to the assumed 
 N$_{\rm H}$ of undetected objects, 
 we also considered the conversion
 factors corresponding to the highest and minimum 
 conversion factors of  
 the X-ray detected 
 Class\,0-I objects.\footnote{We  used the median conversion factor 
  to derive the X-ray luminosity 
(log $L_{t,c}=28.52$\,erg/s)
of the COUP source 1197 for which \citet{getm05a} did not perform
a spectral fit  because of the low counting statistics.} 
We used the same procedure to estimate the upper limits to the X-ray
luminosities in the 2.0--8.0\,keV hard band.

The X-ray properties of the Class\,0-I  COUP X-ray sources, taken from 
\citet{getm05a}, are given in Table\,\ref{coup_cl0I}. We provide
sequential numbers from our catalog, sequential COUP
numbers,  background-corrected counts in the 0.5--8.0\,keV band,
the fractions of the PSF within the source extraction areas, 
effective exposure times,
the hydrogen column densities
 obtained from the spectral fit,  plasma temperatures, 
  emission measures and finally the  absorption-corrected  X-ray 
 luminosities in the 2.0--8.0\,keV hard band, L$_{h,c}$, and 
 in the 0.5--8.0\,keV total band, L$_{t,c}$.

  Values
computed for the X-ray undetected Class\,0-I stars are given in 
Table\,\ref{cl0I_up_lim}. We list
sequential numbers from our catalog, the
celestial coordinates from the {\it Spitzer} catalog, 
upper limits to the source counts computed
with PWDetect in the hard and total bands,
 the effective {\it Chandra} exposure time at the object positions and,
 finally, the upper limits to the unabsorbed X-ray luminosities
 in the hard and total energy bands.

X-ray properties of X-ray detected Class\,II and III stars, 
taken from \citet{getm05a},
are given in Tables \ref{coup_clII_1}   
and \ref{coup_clIII_1} where column names are analogous to those given in
Table\,\ref{coup_cl0I}.

The X-ray luminosities of 
X-ray undetected 
Class\,II and III objects were derived, as for Class\,0-I stars, 
using the median conversion factor 
of the X-ray detected stars in the same class.
This assumption is based on the fact that the NIR colors of the
 X-ray undetected Class\,II stars  indicate
 that  these
 objects are not more absorbed than those detected with COUP.
 In addition, the conversion factors
corresponding to the lowest and highest  observed N$_{\rm H}$ values 
 were also  considered
($10^{21}$ and 4.0 $\times 10^{22}$  cm$^{-2}$  for the Class\,II stars and
 0.5$\times 10^{21}$ and $2.0\times 10^{22}$  cm$^{-2}$  for the Class\,III stars). 
 Values
computed for the X-ray undetected Class\,II and III stars are given in 
Table\,\ref{clII_III_up_lim}, analogous to 
Table\,\ref{cl0I_up_lim}.

\subsection{X-ray luminosity functions\label{xlf_section}}
\begin{figure*}[!ht]
\centerline{\includegraphics[width=18cm]{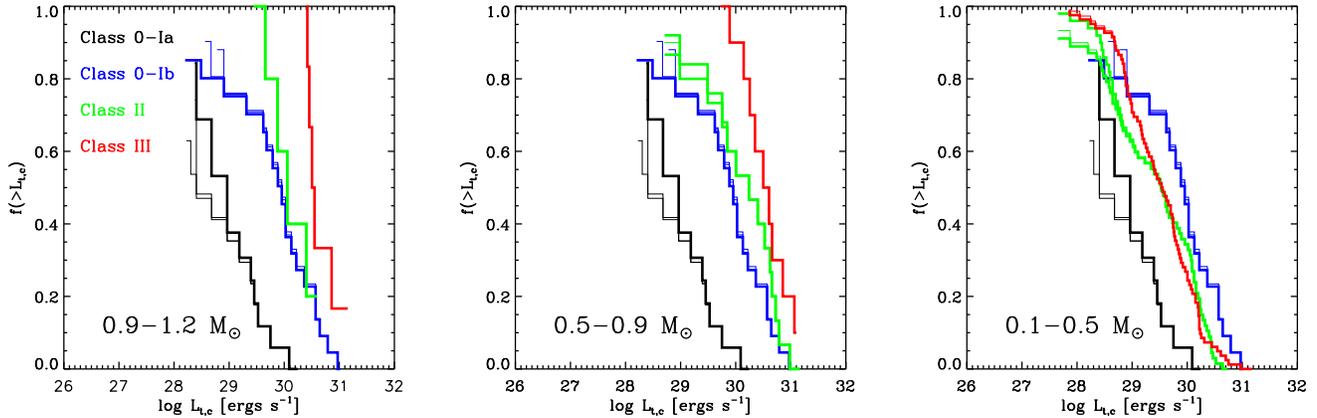}}
\caption{Maximum likelihood cumulative XLF in the total (0.2--8.0)\,keV 
energy band
of the four NIR classes. The three panels refer to Class\,II and Class\,III
stars in three different
 mass ranges. The red and green lines 
indicate the XLF of Class\,III and II objects, respectively, with masses in the
given  range, while  the 
black lines are the XLF of Class\,0-I stars, for which masses are unknown
and are therefore the same in the three panels. Lines  of the same color 
in each panel show the XLF computed using the three different luminosity 
conversion factors for the X-ray non detections. 
Thick lines indicate the XLF used to perform the statistical tests  (see text).
Note that the
distributions do not reach the f=100\% because of the presence of upper limits.}
\label{f08}
\end{figure*}
\begin{figure*}[!ht]
\centerline{\includegraphics[width=18cm]{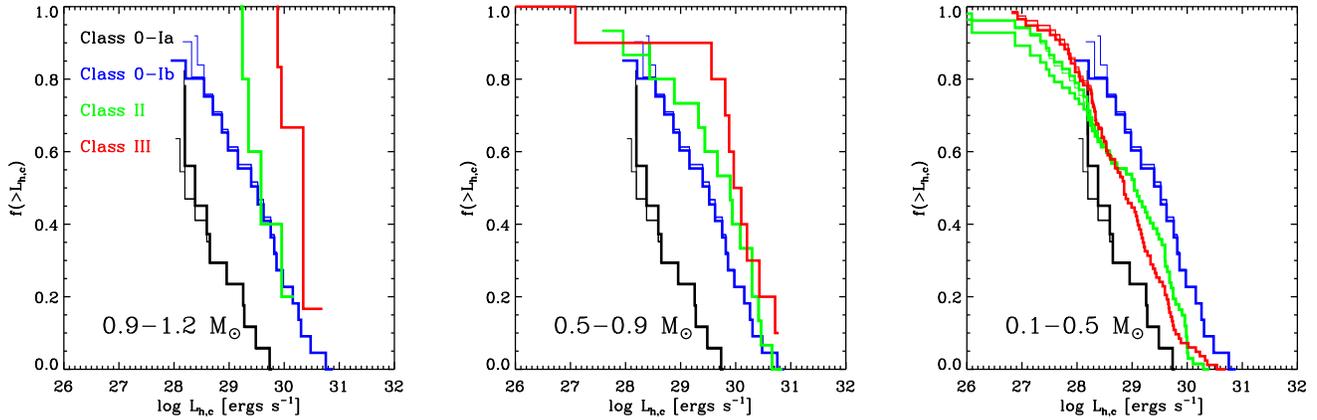}}
\caption{Maximum likelihood cumulative XLF in the hard (2.0--8.0)\,keV 
energy band
of the four NIR classes. The figure is analogous to Fig.\,\ref{f08}.}
\label{f09}
\end{figure*}

We now compare the X-ray luminosity functions  of the stars of the four
considered NIR classes. In order to 
 take into account 
nondetections, XLFs were derived using the
Kaplan-Meier maximum-likelihood estimator.
For Class\,II and III sources, we considered separately stars in the three mass
ranges used by \citet{prei05a}.
Fig.\,\ref{f08} shows the comparison of the  XLFs
of the four populations. The  red and green lines 
indicate the XLF of Class\,III and II objects, respectively, with masses in the
given range, while  the 
 black and  blue lines are the XLFs of Class\,0-Ia and 0-Ib stars,
 for which masses are unknown
and are therefore the same in the three panels. Lines of the same color 
 in each
panel show the XLF obtained 
using the three different count rate to luminosity conversion factors
adopted for the X-ray non detections of each subsample and give an idea of the
uncertainty of the XLFs. 
We removed from this analysis the four objects of 
our sample of Class\,0-Ia protostars  without detections
at 4.5\,$\mu$m. Since this band is
one of the most sensitive bands and the least contaminated by nebulosity,
these objects could be PAH  contaminated sources or even knots of nebulosity. 
By discarding these objects, we are more confident to consider a sample of 
protostar candidates as less contaminated as possible.
The figure shows that 
 for the  two highest mass
ranges,  the X-ray
luminosities of Class\,III stars are highly elevated with respect to those of 
Class\,II stars and 0-Ib objects. These latter two classes have very similar XLFs and
 are significantly brighter than Class\,0-Ia stars. 

  Because the Class\,0-Ia and Class\,0-Ib stars in our sample are heavily
  absorbed, 
  a note of caution is needed. Since absorption precludes the observation
  of the soft part of the intrinsic source X-ray spectra, we 
  cannot rule out the presence of soft, completely absorbed  emission
from Class\,0-I stars. 
Nevertheless,
we know from previous studies of  CTTS stars that a very soft
 component, kT$<$0.3\,keV, in the few cases in which it is observed,
  is mainly due to accretion \citep{kast02,schm05,gude07,argi07,huen07,gros07},
while  harder components are due to coronal-like emission.
This is in agreement with the previous
finding for Class\,II stars in other regions \citep[e.g.][]{stel04},
where the X-ray spectra of Class\,II's were found to include an
additional softer emission component compared to WTT's.  Our
Orion stars are too heavily absorbed to detect this soft component.

Coronal emission, as analyzed with  resolution spectra from {\it 
Charge Coupled Devices}, 
can  
usually be decomposed into softer $\sim$0.8--1.0\,keV and harder 
$\sim$2\,keV components \citep{prei05b}. The soft coronal component, like 
the accretion component, might remain unobserved in heavily extincted sources
such as our Class\,0-I stars and  
thus the X-ray luminosities in the total band can underestimate the true
emission in such objects. 
In addition, magnetic flare components can be strong around 0.5--2.0\,keV,
and dominate the X-ray flux. 
For these reasons, we also considered the comparison of the XLFs of different
classes,
by restricting the analysis to the X-ray
 luminosities in the hard (2.0--8.0)\,keV energy band. 
 We note that we recomputed with PWDetect the upper limits in the hard band for a 
 total of 31 undetected objects, assuming a threshold of 4.8$\sigma$ estimated
from the background in this band. Among these objects, 
 12 are flagged by the PWDetect code
as affected by a X-ray detected source nearby, which therefore raises the upper
limit count rate. In two cases out of 12, however, the asymmetry of the Chandra
PSF at large off-axis angles is such as to make PWDetect clearly overestimate 
the contribution of the detected source to the computed upper limit at the
desired position, which is instead dominated by the background; in these
instances we computed the upper limit from the background alone. 
The results of this analysis are shown in Fig.\,\ref{f09},
analogous to Fig.\,\ref{f08} but for the hard energy band.  

To check if these results are statistically 
significant, we performed statistical tests  developed for ``survival
analysis", suited for the analysis of censored datasets. The
probabilities that the X-ray luminosities of the stars in two samples are drawn
from the same parent distribution were computed with the
ASURV\footnote{Astronomy Survival Analysis  available from the
StatCodes http://www.astro.psu.edu/statcodes} package \citep{feig85}.
 The results of this
analysis are given in Tables\,\ref{xlf_prob} and \ref{xlf_prob_hard} 
where, for each pair of samples,
 we list the probabilities
computed using the five tests of the ASURV package.
 We note that the tests were
performed in the most conservative way, i.e. by considering for each pair
of samples the nearest XLFs among those computed with different conversion factors
for the upper limits. 
\input{tab2}
\input{tab3}
The tests show  that Class\,0-Ia and 0-Ib stars  
have statistically different XLFs both in the hard and total bands, 
at a significance level larger than 99\%;
Class\,0-Ia and II are different 
at a significance level of 99\%  for the [0.9-1.2] and [0.5-0.9]\,M$_\odot$ 
mass ranges. The difference is marginal, both
in the total and hard band,  
for the [0.1-0.5]\,M$_\odot$ 
mass range.
 The difference between Class\,II and III objects
 is  quite marginal (between 85\% and 95\%) for the highest mass range, 
 while for lower mass stars, Class\,II and III stars do not show any
significant difference.

These results  clearly indicate an evolution of the X-ray activity
from Class\,0-Ia, the least X-ray luminous objects, up to Class\,III, 
which show the highest X-ray luminosities, at least for stars with mass larger
than 0.9\,M$_\odot$.  Although   
the difference in the intrinsic
X-ray activity between Class\,II and III stars is quite marginal,
it is in agreement with 
 previous results indicating that the X-ray luminosities of accretors in the ONC are smaller
 than those of non accretors \citep{flac03b,prei05b}. Note, however, that these 
  results cannot be compared directly   because of the different 
  criteria adopted to select the two populations.
 \begin{figure}[!h]
\centerline{\includegraphics[width=9cm]{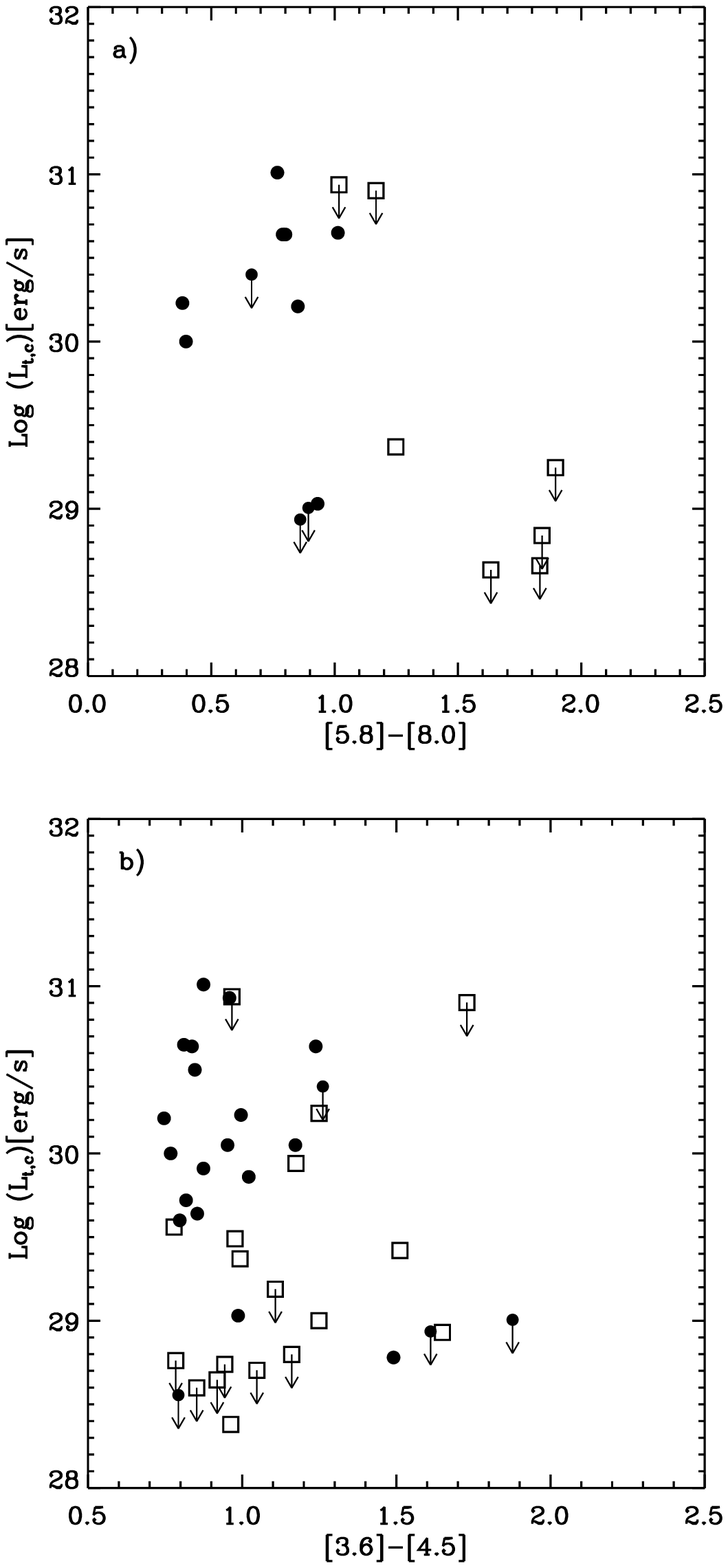}}
\caption{Absorption corrected
X-ray luminosity  of  Class\,0-Ia stars (squares) and 
 Class\,0-Ib (circles) as a function of the [5.8]-[8.0] (panel a) and 
 [3.6]-[4.5] (panel b) colors.
 Upper limits are indicated by arrows.}
\label{f10}
\end{figure}

The most significant and novel result is the difference in X-ray activity level 
between Class\,0-Ia and Class\,II and 
between Class\,0-Ia and Class\,0-Ib sources.
Our results are even more robust if we include the four objects without 
detections
at 4.5\,$\mu$m.

 Based on our selection criterion, Class\,0-Ia objects are protostars with SEDs
increasing up to 8.0\,$\mu$m, while Class\,0-Ib are protostars with SEDs
increasing up to 4.5\,$\mu$m and decreasing at longer wavelengths. 
%
 Fig.\,\ref{f10}a shows the absorption corrected
X-ray luminosities plotted vs. the [5.8]-[8.0] color, that 
discriminates Class\,0-Ia stars (squares) 
from Class\,0-Ib ones (filled circles), the  former
being redder than [5.8]-[8.0]$\simeq$1.1 and the latter bluer than this value.
This plot indicates a trend of increasing X-ray luminosity with decreasing  
[5.8]-[8.0] color. 
Fig.\,\ref{f10}b shows the same luminosities as a function of the [3.6]-[4.5]
color.
 While this latter color does not discriminate between Class\,0-Ia and
0-Ib objects, it clearly confirms that, from the X-ray point of view,
Class\,0-Ia and 0-Ib  objects belong to two different populations. This again
suggests us
that the contamination by reddened Class\,II objects can be very high 
for Class\,0-Ib
objects, while the Class\,0-Ia sample is dominate by true protostars. 
We could then be witnessing the onset
of X-ray activity in very young YSOs.

Class\,0-I objects are affected by larger absorptions with respect to the other
classes as we will confirm in Sect.\,\ref{gas_dust} 
using the X-ray derived
N$_{\rm H}$ values. Can this fact results in a systematic underestimation of 
 L$_{\rm X}$ values for Class 0-I stars? Although we do apply absorption 
corrections  to
the observed X-ray fluxes when deriving luminosities both in the 0.5-8.0\,keV
and 2.0-8.0\,keV bands, these are corrections to the {\em observed} X-ray
spectra and emission from relatively cool plasma would remain unobserved, and
therefore unaccounted for, in heavily observed sources.

In order to verify that the X-ray luminosities of heavily absorbed Class\,0-I
objects are accurately corrected for absorption, we assumed that these sources
have X-ray spectra similar to  Class\,II and III objects, and computed through
extensive simulations the effect of varying the absorbing hydrogen column
density on the determination of X-ray luminosities.

For this purpose we considered the best fit spectra of a subsample of $\sim$200
Class\,II and III objects with small absorption 
(Log\,N$_{\rm H} <21.5$cm$^{-2}$), consisting of either one or two thermal
 components.  For each
of these spectra, using XPSEC and the {\sc fakeit} command, we computed 10
simulated spectra for each of five different  N$_{\rm H}$ values in the range
$10^{21}$cm$^{-2}<$N$_{\rm H} <10^{23}$cm$^{-2}$. We then analyzed the
simulated spectra with a procedure analogous to that used  by \citet{getm05a}
for the analysis of the observed spectra. Each spectrum was rebinned so to have
a minimum of 15 photon per channel and was then fit with both an isothermal and
a two component model. After adopting the simplest of the two models that gave
a statistically acceptable fit, we computed the unabsorbed flux in the
0.5-8.0\,keV and 2.0-8.0\,keV bands. We then computed the absorption corrected
XLFs for each of the five absorption values. Most of the simulated sources did
not have enough photons when absorbed by N$_{\rm H} =10^{23}$cm$^{-2}$. In
order to avoid selection effects, since we are not accounting for upper
limits,  we could only compare the XLFs for the 115 sources whose spectra were
bright enough to be analyzed even when considering the highest absorption
value. The result is that the 5  XLFs for the 0.5-8.0\,keV luminosities have
median within $\sim$0.1\,dex but tend to become wider as the $N_H$ is
increased. The five XLFs for 2.0-8.0\,keV luminosities are instead very similar
in shape and with medians within 0.03\,dex, except for the one relative to the
highest absorption values which is {\em brighter} by $\sim$0.1 dex. We conclude
that, in the assumption that Class 0-I sources have the same spectra of Class
II-III ones, the X-ray luminosities we derive are not underestimated.

We cannot 
exclude that some of the
  Class\,0-Ib stars are actually heavily absorbed Class\,II stars
  (see Fig.\,\ref{f03}).  However,
even in this case, 
our result would 
  plausibly indicate a significant
  evolution of X-ray activity in the early stages of YSO formation.
  
Due to various IR sensitivity limitations at very high
obscurations, the Class\,0-I sampling  might be restricted to higher masses
(with stronger IR bolometric luminosities), and through the L$_{\rm X}$-mass
correlation, would have higher L$_{\rm X}$ values introducing a bias. But we find
that these heavily obscured stars have lower, not higher, L$_{\rm X}$ values, further
supporting our result.

Another bias may be introduced by the requirement that all stars considered have
IRAC detections in one or more bands. Therefore, it is possible that lower-mass
(i.e. bolometrically weaker) Class\,III stars have been systematically omitted
from the sample, leaving behind too many higher mass, i.e. stronger L$_{\rm X}$ stars.
But this is not relevant as the XLF comparisons in Figs.\,\ref{f08}
and \ref{f09}
are binned by mass. We might have few 0.1--0.5\,M$_\odot$ Class\,III stars, but
those shown should be  unbiased in L$_{\rm X}$.

%
%

Another possible worry is that flares could dominate the flux from some of our 
objects, and our X-ray spectra could pertain to the flaring state rather than
the quiescent state.  We have verified  that this is not the case by using the
technique described in \citet{wolk05} to recompute X-ray spectra and fluxes for
our stars using photons detected only during non-flaring intervals. Fits to
these new spectra yield XLFs that do not differ significantly  from those shown
in Fig.\,\ref{f08}. We have also verified that using these new spectra
the results discussed in the next section do not change significantly.

We note that our results are consistent with the non-detection of class\,0
sources of Serpens \citep{giar07}. In fact, their limiting sensitivity, 
(4.0$\times 10^{29}$\,erg/s in the hard band) estimated
assuming temperature and absorption similar to those we found for our X-ray
detected Class\,0-I sources, corresponds to the highest tenth percentile of the
XLF of our Class\,0-Ia stars. Considering the small size of the Serpens 
Class\,0 star sample, their non-detection is not surprising.
\subsection{Gas and dust absorption\label{gas_dust}}
\begin{figure}[!h] 
\centerline{\includegraphics[width=9cm]{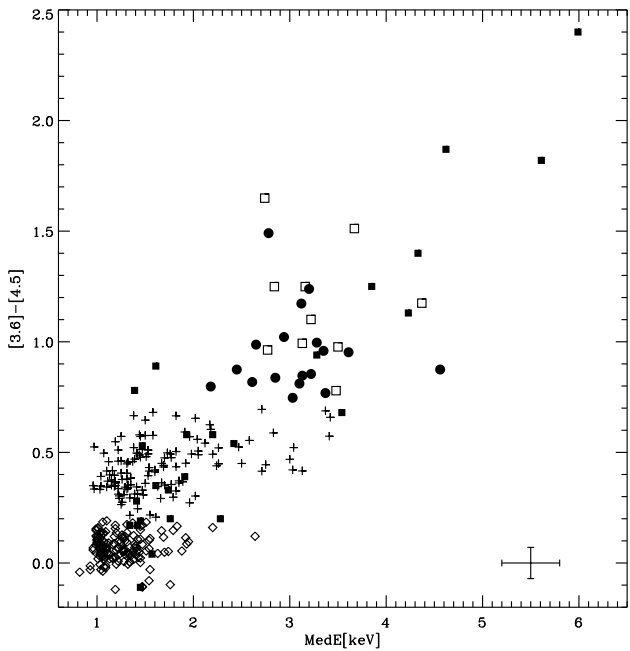}}
\caption{{\it Spitzer}/IRAC [3.6]-[4.5] color vs. {\it Chandra} X-ray median
energy of all selected sources. The different classes ONC objects
are indicated using the same symbols of Fig.\,\ref{f05}, while filled
squares are the values of the YSO in IC\,1396N taken from \citet{getm07}.
Typical error bars in the median energy 
and in the IRAC colors are also shown.} 
\label{f11}
\end{figure}

Figure\,\ref{f11} shows the
 {\it Spitzer}/IRAC [3.6]-[4.5] color vs. {\it Chandra} X-ray median
energy of all selected sources. The typical uncertainty  in the median energy 
has been estimated using the results of the MARX simulations made in 
\citet{getm06},
where the median energy errors were computed as a function of the source net 
counts. We used the value of 0.3\,keV, corresponding to objects with 
MedE$\gtrsim$2\,keV and with about 30 net counts which is  the 
minimum of net counts we have for Class\,0-I objects. The typical 
uncertainty  in the IRAC
color  has been estimated assuming a 5\% uncertainty for the IRAC magnitudes.

As discussed in \citet{getm07}, high X-ray median energy 
(MedE$\gtrsim1.7$\,keV) can be considered as
an indicator of absorbing column density that arises from gas,
primarily metallic elements
 (e.g. N, O, Ne, Mg, Si, Ar, Fe, etc.). On the other hand, 
if the object is known to be a protostar, the [3.6]-[4.5] color
is primarily  a measure of the density of grains in the local infalling envelope.
Thus, the correlation between these two quantities, first found in 
\citet{getm07}, has been interpreted by these authors as evidence of absorption
from the protostellar envelope rather than from the 
ambient molecular cloud material. 
 Although our data do not cover the whole  X-ray
median energy, they show a trend  consistent with that found using the IC\,1396N
data, also shown in Fig.\,\ref{f11}.  Nevertheless, the 
interpretation of our data is not as clear as it was in IC1396N,
since the ONC background is quite complex while
 the maximum line-of-sight column is A$_V\sim10$\,mag.

\begin{figure*} 
\centerline{\includegraphics[width=16cm]{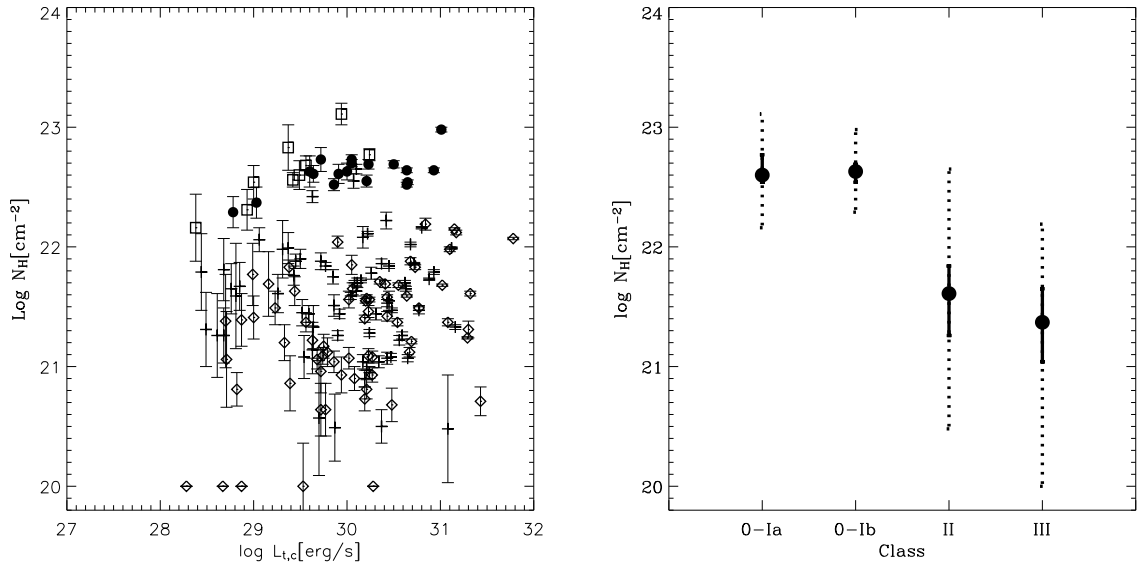}}
\caption{Left panel shows the log\,N$_{\rm H}$  
values as a function of the unabsorbed X-ray
luminosity L$_{t,c}$ for the X-ray detected objects. 
Symbols are as in Fig.\,\ref{f05}.
Right panel shows, for each
class, the median values of the absorbing column density distributions
indicated by the big black dots.   
Solid bars indicate the values of log\,N$_{\rm H}$
included between the first and third quartiles, 
while the dashed lines (partially hidden by solid bars) 
indicate the values included between the minimum and
the maximum of the distributions of the 
log\,N$_{\rm H}$ values.
 Note that only objects with errors in log\,N$_{\rm H}$ smaller
than 0.5 were considered.} 
\label{f12}
\end{figure*}

Fig.\,\ref{f12}    
 compares the absorbing column density N$_{\rm H}$
for the  different populations. 
In the left panel, log\,N$_{\rm H}$  
is plotted for the X-ray detected objects as a function of the unabsorbed X-ray
luminosity L$_{t,c}$.
The right panel shows, for each
class, the median values of the absorbing column density distributions,
values between the first and third quartiles and 
values between the minimum and the maximum of the distributions of the 
log\,N$_{\rm H}$ values. Note that only objects with   
errors in log\,N$_{\rm H}$ smaller than 0.5 have been considered. However, the
median values of the distribution do not change significantly if all data are
considered. 

A clear trend of increasing absorbing column density from Class\,III to
 Class\,0-I can be seen both  in the left and 
right panels. This trend is
expected since Class\,0-I objects are characterized by a large amount of 
infalling material and are also typically embedded in the molecular cloud; 
a smaller amount of obscuring material is instead present for Class\,II
objects that have shed their envelope and 
formed a circumstellar disk, while the extinction suffered
from Class\,III objects is   due to  interstellar material only
(both within the cloud  and on the line of sight to the cloud). 
 %

We  note that the estimate of the N$_{\rm H}$  values of Class\,0-I
stars is based primarily on 
the hard component of the
 X-ray spectrum.   That is, at the temperatures of the protostars 
 (kT$_{\rm av}\sim2$\,keV)
  and with log N$_{\rm H}\sim23$\,cm$^{-2}$,
 the observed flux is almost entirely due to  the hard component 
 of the X-ray spectrum.
 At the same temperatures, if the N$_{\rm H}$  was
 smaller, the total flux
 would be distributed  between the  soft and hard component and the shape of
 the spectrum would be different.
 Our technique should thus be the most accurate for objects with high
  N$_{\rm H}$ values -- thus yielding particularly good  N$_{\rm H}$ values
  for the, on average, heavily embedded protostars.
 Therefore, we argue that the low XLFs for protostars we have derived
 (see Figure 9) are unlikely to be due to errors in our estimates
 of the X-ray absorption towards these objects, as confirmed from the analysis
 of the hard XLFs and the simulated spectra of Class\,II and III (see
 Sect.\,\ref{xlf_section}).

%
\subsection{Evolution of YSO X-ray spectra and temporal characteristics}
\begin{figure*} 
\centerline{\includegraphics[width=16cm]{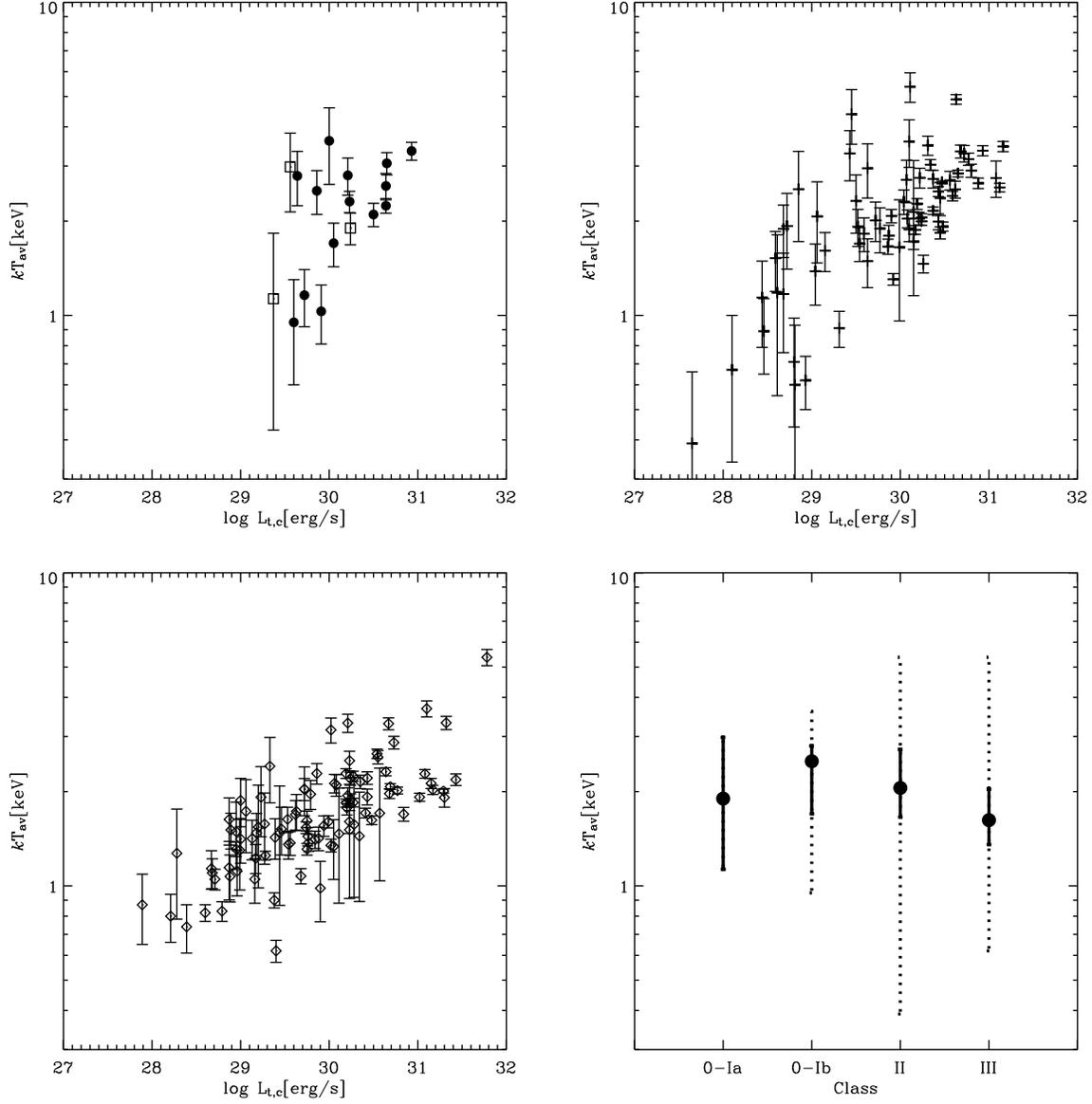}}
\caption{ kT$_{\rm av}$
values as a function of the unabsorbed X-ray
luminosity L$_{t,c}$ for the X-ray detected objects, 
 plotted separately for Class\,0-I, Class\,II and Class\,III stars,
  in the first three panels.
Symbols are as in Fig.\,\ref{f05}.
The bottom-right panel shows, for each
class, the median values of  the average plasma temperature distributions
 indicated by the
big black dots. 
Solid bars indicate values of the temperature distributions between 
the first and third quartiles, while  dashed lines  (partially hidden by 
solid bars)
 indicate values between the minimum and the maximum 
 of the distributions of the
 kT$_{\rm av}$ values. 
 Note that only objects with errors 
in kT$_{\rm av}$ smaller than 1\,keV were considered.} 
\label{f13}
\end{figure*}

Fig.\,\ref{f13} shows 
the results of the comparison  of the average 
plasma temperature $k$T$_{\rm av}$ amongst the
different classes.  For X-ray spectra fit with two isothermal
components, $k$T$_{\rm av}$ was 
computed as the mean of the soft and hard $k$T spectral components,
weighted by their respective emission  measures; 
for spectra fit with a single component $k$T$_{\rm av}$=$k$T.
The  $k$T$_{\rm av}$ values for the X-ray detected objects
of the different populations
are plotted as a function of the unabsorbed X-ray luminosity L$_{t,c}$.
 Note that only objects with  
errors on $k$T$_{\rm av}$ smaller than 1\,keV have been considered.
However the
 distributions do not change significantly if all data are
considered.  
 The bottom-right panel of Fig.\,\ref{f13} shows, for each
class, the median values of the average plasma temperatures, 
values between 
the first and third quartiles and values between the minimum and the 
maximum of the distributions of the 
$k$T$_{\rm av}$ values. No trend in plasma energy is found as a function of
evolutionary stage from Class\,0-Ia to Class\,III systems.

A trend of increasing of the average plasma temperature with 
increasing unabsorbed
X-ray luminosity is clearly evident for all the samples, as has often been
reported for other types of coronal sources
\citep[e.g.][]{schm90}. A similar result is obtained 
by considering only the subsample of stars with X-ray spectra fit with 
two temperature components:
high luminosity stars tend to have high values of the ratio
between the high and low temperature emission measures (EM$_2$/EM$_1\sim 2.5$),
 suggesting that the hot plasma component dominates over the cold one
for  the high luminosity stars. 
No significant difference between the  classes is observed in 
Fig.\,\ref{f13}.

Finally we checked if ONC stars of different classes  show different
  X-ray temporal behavior. To do this, we used the results of
  the nonparametric one-sample Kolmogorov-Smirnov (K-S) test performed by
  \citet{getm05a}. The results of the K-S test, P$_{\rm KS}$, strongly depend on the source
  count-statistics and therefore on their count rate. 
  For this
  reason, we compared the probabilities of variability P$_{\rm var}$=(1-P$_{\rm KS}$)
  for the  different
  classes of stars with count rates in definite ranges. 
 The results are given in Table\,\ref{probks_var} where, for each
class and for a given range of Count Rate (CR), we report  the number of stars with 
P$_{\rm var}>99$\%  with respect to the total number of stars of each
class in the corresponding range of count rates. The percentage of variable
stars is also indicated within brackets. Even if the statistics for each sample
is quite poor, the percentages indicate that the  different classes  
have very similar fractions of variable stars.
In practice, when observed with high statistics, all the sources are
variable, independently  on their class.
\input{tab4}
%
%
\section{CLASS\,0-I: COUNTERPARTS AT OTHER WAVELENGTHS\label{discussion}}
We list in Table\,\ref{classI_matches}  the counterparts of the  Class\,0-I
objects identified in this work that have been detected in 
X-rays (the COUP survey), JHKL bands
 \citep[][ hereafter LMH00 and MLL02, respectively]{lada00,muen02},
10 and 20\,$\mu$m \citep[][ hereafter RBP5]{robb05a} and in the 
11.7\,$\mu$m  \citep[][ hereafter SBS05]{smit05}.
None of our candidate Class\,0-I sources
are in the list of 1.3\,cm radio sources of \citet{zapa04b}.

The JHKL photometric survey from \citet{lada00}  included 391 stars
 in the central 6\Min55$\times$6\Min55
of the Trapezium cluster. By selecting stars with $K-L>$1.5, they obtained a list
of 78 candidate protostars. We correlated this list with our catalog using a matching
radius of 1\Sec2. As expected, the \citet{lada00}   sample, selected on the basis of
the $K-L$ color, is larger
than our sample for which the selection requires at least two colors computed using the K 
and the four IRAC magnitudes. For this
reason, we find that only 4  of the 78 \citet{lada00} candidate protostars 
are in common with our Class\,0-Ia stars and 9 with our Class\,0-Ib stars.
 Among the remaining 65 objects, 16 cannot be
Class\,0-I objects since they have an
optical counterpart in our catalog and 27 have
[3.6]-[4.5]$<$0.7 and are therefore consistent with star-disk systems (Class\,II)
rather than protostars; for the remaining 22 \citet{lada00} candidate protostars,
the available magnitudes do not
satisfy our selection criteria.

Among our Class\,0-I objects, only 5 (Class\,0-Ib) have been detected at 
 10 and 20\,$\mu$m by \citet{robb05a}; of these 5, 2 are in the list of 
the   embedded objects previously studied in \citet[][ hereafter SBS04]{smit04}  
in the OMC-1S region  at  8.8, 11.7 and 18.8\,$\mu$m.

\begin{figure*} 
\centerline{\includegraphics[width=16cm]{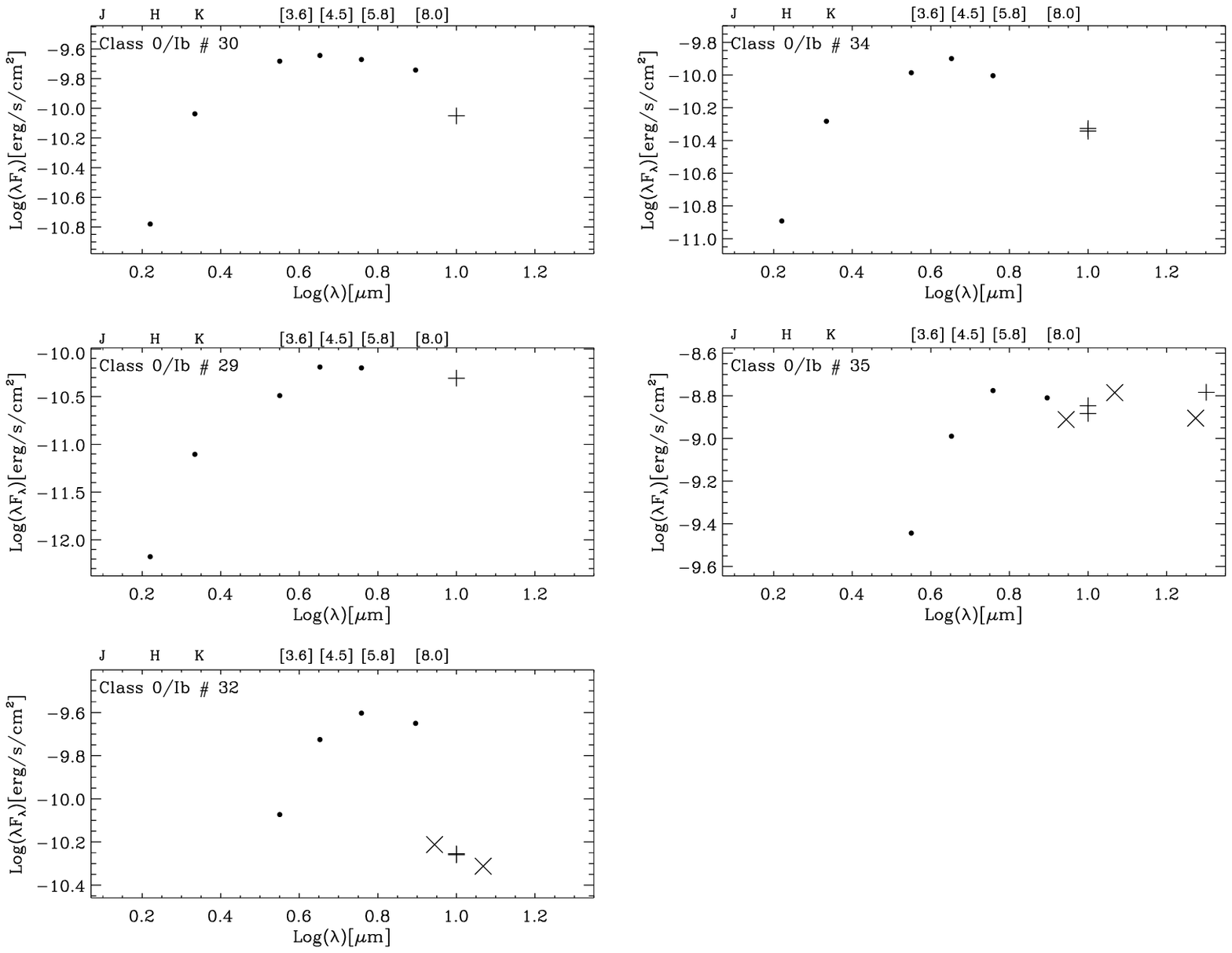}}
\caption{SED of Class\,0-I candidate objects in the \citet{robb05a} list
with photometry at 10 and/or 20\,$\mu$m. Dots indicate fluxes computed
used the magnitudes from our catalog,   plus symbols correspond to
 the magnitudes at 10 and 20\,$\mu$m from the \citet{robb05a} catalog and
 {\tt X} correspond to the magnitudes at 8.8, 11.7 and 18.8\,$\mu$m from the
 \citet{smit04} catalog. Multiple values at 10\,$\mu$m are 
 relative to the two different nights of observation in \citet{robb05a}.}
\label{f14}
\end{figure*}

We interpolated the flux conversion coefficients given in 
\citet{lada06} to  10 and 20\,$\mu$m and we computed the corresponding
fluxes. SEDs with all the available fluxes for these 5 objects are shown in 
Fig.\,\ref{f14}.  

In order to understand the properties of our sample of Class\,0-I objects,
we also considered the compact radio sources detected by \citet{zapa04b} and 
\citet{zapa04a} 
using, respectively,  1.3\,cm (in a region of 30\sec$\times$30\sec around
 OMC-1S) and 3.6\,cm observations 
 (in a region of 4\min$\times$4\min\ around the Trapezium cluster) 
 performed with the Very Large Array. 
With a matching radius of
1\sec, we retrieved in our catalog  7 of the 11 radio sources at  1.3\,cm
and 67 of the 77 compact     3.6\,cm  sources. 

None of the \citet{zapa04a,zapa04b}  matched sources is in our list of Class\,0-I 
objects. The absence of 1\,cm radio counterparts in our protostar sample
might imply that 
the radio sources are mainly magnetically active non-thermal Class\,III stars
rather than thermal protostellar outflows.

Note that the north-east region of the COUP FOV, where most of our candidate
 Class\,0-I are concentrated, is not included in the FOV of the literature sources
 mentioned in this section. For this reason, the SEDs of Class\,0-I objects in
 this region cannot yet be extended to wavelengths longer than 8\,$\mu$m, which
 is crucial to confirm (or refute) their protostellar nature.

Finally, our list of candidate Class\,0-I protostars includes only two objects
(COUP sources 420 and 484) of the list of stars  studied in \citet{gros05}, 
located at the bottom edge of the OMC-1S region, i.e., far from
the density peak and its associated nebular emission.
 We know that the absence of 
IRAC protostars near 
OMC-1s and in the BN/KL region is due to nebular contamination and crowding, 
not to the real absence of protostars in these  two high density regions. In
addition, we cannot classify
 the COUP sources 582, 594, 615, 633, 641, 659 and 667, indicated in
 \citet{gros05} as embedded YSOs in a subcluster, since they do not have any
 counterpart    at the IRAC bands.
 

%
\section{SUMMARY AND CONCLUSIONS}

We have selected a sample of 45 candidate
 Class\,0-I objects in the ONC field observed in the X-rays by COUP
using new deep $JHK$, [3.6], [4.5], [5.8] and [8.0] photometry obtained at
the ISPI@4m CTIO telescope and from {\it Spitzer} IRAC observations. New
deep optical observations taken with the WFPC2 camera of HST and the WFI camera 
of ESO\,2.2\,m
were also used to reject candidate Class\,0-I objects with optical counterparts.

We distinguish between candidate protostars 
with SEDs increasing monotonically from K up to 8\,$\mu$m, indicated as 
Class\,0-Ia, and 
candidate protostars with SED increasing from K up to 4.5\,$\mu$m and 
decreasing at longer
 wavelengths, indicated as Class\,0-Ib, since  the different slope of the SED
 at wavelengths larger than 4.5\,$\mu$m could be an
indicator of a different density in the circumstellar material. 
 
Out of the 23 Class\,0-Ia stars, 10 have been detected in the X-rays with COUP,
 while out the 22 Class\,0-Ib stars, 18 have been  detected; 
we are thus in the position to study, for the first time,
 the X-ray properties of a significant and homogeneous sample of  candidate
protostars. 

Samples of bona fide more evolved 
Class\,II and III stars in the same region
  have also been selected in order to study the time and mass evolution of
    X-ray activity.

Our principal result is that
Class\,0-Ia stars are significantly less luminous in X-rays, both in the total
and hard bands, than more
evolved Class\,II stars with mass larger than 0.5\,M$_\odot$, 
while Class\,0-Ib stars show X-ray luminosities similar
to Class\,II stars; finally, our data confirm previous determinations 
\citep{neuh97,flac03b,prei05b} that 
 Class\,III stars with masses 
$>$0.3\,M$_\odot$
 are more X-ray luminous than Class\,II objects, even if the result
is quite marginal with our data.
  
Our result that Class\,0-Ia objects are less luminous in X-rays than objects of
other classes, support the hypothesis that the onset of X-ray emission occurs at
a very early stage, when the objects show an increasing  SED at least up to 
8.0\,$\mu$m. 
The lack of a detailed theory of protostellar evolution does not
allow us to interpret the  difference in the evolutionary stage
 between Class\,0-Ia and 0-Ib protostars.
 Although Class\,0-Ib objects are selected as candidate protostars, our data
suggest that this sample could be dominated by reddened Class\,II stars.  
 If we consider the  X-ray limiting sensitivity in \citet{giar07}, we deduce
  that our results are in agreement with the non-detection of
 Class\,0 protostars in Serpens. 
  
The X-ray spectral properties of Class\,0-Ia and 0-Ib 
are similar to those of the more evolved
Class\,II and III objects except for a larger
absorption likely due to the enhanced local density surrounding protostellar
objects.
The  different evolutionary classes also show  similar X-ray variability 
characteristics and the protostellar variability is indistinguishable from
T\,Tauri variability. This result supports the XEST determination that
variability of Class\,II and III stars are the same \citep{stel07} and the COUP
determination that T\,Tauri M stars have the same variability as more massive
stars \citep{cara07}.
Also related is the COUP result that YSOs with high X-ray luminosities have the
same first ionization potential-related elemental abundances as older flaring
stars \citep{magg07}.

 All of these results support the general theme that there
is only one mechanism in YSO X-ray production at energies $\gtrsim$1\,keV, 
and that this mechanism arises
from stellar magnetic activity which does not depend on accretion or disks.
The various recent arguments 
\citep{kast02,schm05,gude07,argi07,huen07,gros07} that accretion accounts for some
X-ray emission and line ratios, probably only refers to soft components which
are not accessible to our study \citep{feig07}.
\acknowledgments
This work is based on observations made with {\it Hubble} Space Telescope and 
the {\it Spitzer} Space Telescope,
which is operated by the Jet Propulsion Laboratory (JPL), California Institute
of Technology under a contract with NASA.
Partial support for this work was provided by NASA through an award issued
by JPL/Caltech, by ASI and INAF. EDF is supported by NASA contract NAS8-38252.
This research has made use of NASA's Astrophysics Data System (ADS)
Abstract Service, and of the SIMBAD database, operated at CDS,
Strasbourg, France.
This research has also made use of data products from the Two Micron
All-Sky Survey (2MASS), which is a joint project of the University of
Massachusetts and the Infrared Processing and Analysis Center, funded by
the National Aeronautics and Space Administration and the National Science
Foundation.  These data were served by the NASA/IPAC Infrared Science
Archive, which is operated by the Jet Propulsion Laboratory, California
Institute of Technology, under contract with the National Aeronautics and
Space Administration.
We wish to thank Francesco Damiani and Konstantin Getman for helpfull
discussions. We also thank the anonymous referee for suggestions that helped
improve this paper. 
\bibliographystyle{aa} 
\bibliography{orion}

\input{tab5}
\input{tab6}
\input{tab7}
\input{tab8}

%
\input{tab9}
\input{tab10}
\input{tab11}
\input{tab12}
\input{tab13}
\input{tab14}

\end{document}

%% file: tab2.tex
\tabcolsep 0.15truecm
\begin{table*}
\caption{Probabilities of correlation between  the XLFs in the total band (0.2--8.0)\,keV of two samples,
computed with the ASURV package (see Section\,\ref{xlf_section}). 
$P_1$ and $P_2$ are the probabilities computed using
the Gehan generalized Wilcoxon test with permutation and hypergeometric
variance, respectively, $P_3$ is the probability from the Logrank test
 while $P_4$ and $P_5$ are the probabilities from the Peto and Peto
and from the Peto and Prentice generalized Wilcoxon tests, respectively.}
\label{xlf_prob}
\centering
\begin{tabular}{cccccccc} 
\hline\hline
             Mass    &    P$_1$    &    P$_2$    &    P$_3$    &    P$_4$    &    P$_5$    &  Sample 1   &  Sample 2  \\
          
       [M$_\odot$]   &      [\%]   &      [\%]   &      [\%]   &      [\%]   &      [\%]   &             &            \\
\hline
    
          --           &    0.23   &    0.21   &    3.13   &    0.36   &    0.25   &Class 0-Ia   &Class 0-Ib  \\
    
 [     0.9--     1.2  ]&    0.35   &    0.07   &    0.03   &    0.46   &    0.45   &Class 0-Ia   &  Class II  \\
    
 [     0.9--     1.2  ]&   94.75   &   94.99   &   62.55   &   97.88   &   97.73   &Class 0-Ib   &  Class II  \\
    
 [     0.9--     1.2  ]&    6.79   &    5.10   &   16.15   &   16.15   &      --   &  Class II   & Class III  \\
    
 [     0.5--     0.9  ]&    0.04   &    0.02   &    0.06   &    0.05   &    0.03   &Class 0-Ia   &  Class II  \\
    
 [     0.5--     0.9  ]&   35.94   &   35.60   &   58.58   &   36.77   &   36.64   &Class 0-Ib   &  Class II  \\
    
 [     0.5--     0.9  ]&   33.13   &   32.21   &   14.63   &   31.51   &   32.68   &  Class II   & Class III  \\
    
 [     0.1--     0.5  ]&    5.28   &    5.64   &   23.77   &    6.04   &    5.53   &Class 0-Ia   &  Class II  \\
    
 [     0.1--     0.5  ]&    3.41   &    2.02   &    1.52   &    3.43   &    2.88   &Class 0-Ib   &  Class II  \\
    
 [     0.1--     0.5  ]&   86.08   &   86.09   &   40.72   &   80.37   &   80.34   &  Class II   & Class III  \\
\hline
\end{tabular}
\end{table*}

%% file: tab3.tex
\tabcolsep 0.15truecm
\begin{table*}
\caption{Probabilities of correlation between  the XLFs  in the hard band (2.0--8.0)\,keV  of two samples,
computed with the ASURV package (see Section\,\ref{xlf_section}). 
The probabilities are those described in Table\,\ref{xlf_prob}.}
\label{xlf_prob_hard}
\centering
\begin{tabular}{cccccccc} 
\hline\hline
             Mass    &    P$_1$    &    P$_2$    &    P$_3$    &    P$_4$    &    P$_5$    &  Sample 1   &  Sample 2  \\
          
       [M$_\odot$]   &      [\%]   &      [\%]   &      [\%]   &      [\%]   &      [\%]   &             &            \\
\hline
    
          --           &    0.36   &    0.36   &    2.99   &    0.53   &    0.47   &Class 0-Ia   &Class 0-Ib  \\
    
 [     0.9--     1.2  ]&    0.57   &    0.15   &    0.08   &    0.72   &    0.73   &Class 0-Ia   &  Class II  \\
    
 [     0.9--     1.2  ]&   94.84   &   95.04   &   46.15   &   93.70   &   94.39   &Class 0-Ib   &  Class II  \\
    
 [     0.9--     1.2  ]&    6.54   &    4.77   &    9.37   &    9.37   &      --   &  Class II   & Class III  \\
    
 [     0.5--     0.9  ]&    0.08   &    0.05   &    0.89   &    0.12   &    0.08   &Class 0-Ia   &  Class II  \\
    
 [     0.5--     0.9  ]&   28.18   &   27.57   &   56.83   &   29.37   &   28.97   &Class 0-Ib   &  Class II  \\
    
 [     0.5--     0.9  ]&   62.80   &   62.59   &   34.13   &   61.58   &   62.47   &  Class II   & Class III  \\
    
 [     0.1--     0.5  ]&   14.23   &   14.57   &   45.40   &   16.79   &   16.58   &Class 0-Ia   &  Class II  \\
    
 [     0.1--     0.5  ]&    3.62   &    2.19   &    2.05   &    3.70   &    3.02   &Class 0-Ib   &  Class II  \\
    
 [     0.1--     0.5  ]&   46.63   &   46.63   &   91.93   &   50.19   &   50.23   &  Class II   & Class III  \\
\hline
\end{tabular}
\end{table*}

%% file: tab4.tex
\tabcolsep 0.15truecm
\begin{table*}
\caption{X-ray variable sources of different evolutionary 
 class stars in the ONC with $P_{\rm KS}>99$\%.}
\label{probks_var}
\centering
\begin{tabular}{ccccc} 
\hline\hline
          log (CR)   &        Class 0/Ia   &        Class 0/Ib   &         Class II    &         Class III  \\
\hline 
  [-5,-4  ]   &   2/ 4 ( 50\%)   &   0/ 2 (  0\%)   &   1/ 4 ( 25\%)   &   0/ 2 (  0\%)  \\
          
  [-4,-3  ]   &   5/ 6 ( 83\%)   &   6/ 9 ( 66\%)   &  15/28 ( 53\%)   &  19/30 ( 63\%)  \\
          
  [-3,-2  ]   &--                &   7/ 7 (100\%)   &  38/38 (100\%)   &  46/50 ( 92\%)  \\
          
\hline
\end{tabular}
\end{table*}

%% file: tab5.tex
\begin{landscape}
\tabcolsep 0.06truecm
\begin{table}
\caption{NIR photometry of Class 0/Ia candidates.}
\label{cl0I_1}
\centering
\begin{tabular}{ccccccccccc} 
\hline\hline
            Seq.   &            COUP   &        RA(2000)   &       Dec(2000)   &       J   &       H   &       K   &   [3.6]   &   [4.5]   &   [5.8]   &           [8.0]  \\
            Num.   &              ID   &           [deg]   &           [deg]   &           &           &           &           &           &           &                  \\
\hline
     1    &            &      83.7093162    &      -5.3641575    &                         &                         &                         &  13.014$\pm$   0.035    &                         &  10.132$\pm$   0.035    &   8.291$\pm$   0.035   \\
     2    &            &      83.7313619    &      -5.2982484    &                         &                         &                         &  13.274$\pm$   0.023    &                         &  10.300$\pm$   0.027    &   8.404$\pm$   0.026   \\
     3    &     274    &      83.7777349    &      -5.3789530    &                         &                         &                         &  12.110$\pm$   0.016    &  10.598$\pm$   0.011    &   9.607$\pm$   0.017    &                        \\
     4    &            &      83.7940632    &      -5.5850858    &                         &                         &                         &  12.866$\pm$   0.017    &  11.137$\pm$   0.004    &  10.008$\pm$   0.024    &   8.840$\pm$   0.012   \\
     5    &            &      83.7956871    &      -5.3627791    &                         &  14.957$\pm$   0.044    &  12.585$\pm$   0.031    &  10.526$\pm$   0.112    &   9.742$\pm$   0.029    &                         &                        \\
     6    &            &      83.8000978    &      -5.3113336    &                         &  16.336$\pm$   0.023    &  14.827$\pm$   0.132    &  11.844$\pm$   0.014    &  10.796$\pm$   0.007    &   9.787$\pm$   0.027    &                        \\
     7    &     484    &      83.8020643    &      -5.4105966    &  15.875$\pm$   0.084    &  15.340$\pm$   0.073    &  13.177$\pm$   0.026    &  10.877$\pm$   0.015    &   9.628$\pm$   0.011    &                         &                        \\
     8    &            &      83.8111246    &      -5.2812258    &                         &                         &  15.111$\pm$   0.017    &  12.590$\pm$   0.018    &  11.482$\pm$   0.031    &                         &                        \\
     9    &     696    &      83.8137692    &      -5.3580026    &                         &  14.500$\pm$   0.010    &  12.203$\pm$   0.011    &  10.066$\pm$   0.020    &   9.103$\pm$   0.014    &                         &                        \\
    10    &     702    &      83.8141265    &      -5.3262308    &                         &                         &                         &  12.004$\pm$   0.015    &  10.355$\pm$   0.009    &   9.240$\pm$   0.017    &                        \\
    11    &     860    &      83.8211125    &      -5.3168295    &                         &  15.237$\pm$   0.010    &  12.850$\pm$   0.009    &  10.468$\pm$   0.011    &   9.491$\pm$   0.008    &                         &                        \\
    12    &     859    &      83.8212087    &      -5.3038593    &                         &                         &  14.637$\pm$   0.012    &  11.509$\pm$   0.021    &  10.260$\pm$   0.007    &   9.406$\pm$   0.047    &                        \\
    13    &            &      83.8255845    &      -5.3414585    &  15.114$\pm$   0.080    &  14.999$\pm$   0.016    &  12.804$\pm$   0.008    &  11.057$\pm$   0.033    &  10.204$\pm$   0.029    &                         &                        \\
    14    &            &      83.8302403    &      -5.3200782    &                         &  16.206$\pm$   0.014    &  13.911$\pm$   0.009    &  12.070$\pm$   0.043    &  11.151$\pm$   0.029    &                         &                        \\
    15    &            &      83.8319297    &      -5.2961715    &                         &                         &                         &  11.352$\pm$   0.037    &  10.191$\pm$   0.010    &   8.963$\pm$   0.048    &                        \\
    16    &    1048    &      83.8320314    &      -5.3249635    &                         &  14.869$\pm$   0.009    &  12.378$\pm$   0.011    &  10.116$\pm$   0.011    &   9.337$\pm$   0.006    &                         &                        \\
    17    &    1115    &      83.8362894    &      -5.3239664    &                         &  15.173$\pm$   0.020    &  11.749$\pm$   0.015    &   8.577$\pm$   0.005    &   7.584$\pm$   0.003    &   6.545$\pm$   0.006    &   5.297$\pm$   0.027   \\
    18    &            &      83.8363643    &      -5.2232223    &                         &                         &                         &  11.194$\pm$   0.009    &  10.227$\pm$   0.023    &   9.405$\pm$   0.024    &   8.388$\pm$   0.035   \\
    19    &    1197    &      83.8425168    &      -5.3355986    &                         &                         &  15.253$\pm$   0.047    &  11.508$\pm$   0.020    &  10.407$\pm$   0.011    &                         &                        \\
    20    &            &      83.8479858    &      -5.4635150    &                         &                         &                         &  13.088$\pm$   0.039    &                         &  10.216$\pm$   0.044    &   8.384$\pm$   0.070   \\
    21    &    1321    &      83.8554596    &      -5.3351016    &                         &                         &                         &  11.293$\pm$   0.022    &  10.118$\pm$   0.012    &   8.572$\pm$   0.116    &                        \\
    22    &            &      83.8634267    &      -5.4475646    &                         &                         &                         &  11.977$\pm$   0.018    &                         &   9.206$\pm$   0.039    &   7.573$\pm$   0.045   \\
    23    &            &      83.8997573    &      -5.3532412    &                         &  16.440$\pm$   0.016    &  14.683$\pm$   0.011    &  12.755$\pm$   0.028    &  11.812$\pm$   0.023    &                         &                        \\
\hline
\end{tabular}
\end{table}
\end{landscape}

%% file: tab6.tex
\begin{landscape}
\tabcolsep 0.06truecm
\begin{table}
\caption{NIR photometry of Class 0/Ib candidates.}
\label{cl0I_2}
\centering
\begin{tabular}{ccccccccccc} 
\hline\hline
            Seq.   &            COUP   &        RA(2000)   &       Dec(2000)   &       J   &       H   &       K   &   [3.6]   &   [4.5]   &   [5.8]   &           [8.0]  \\
            Num.   &              ID   &           [deg]   &           [deg]   &           &           &           &           &           &           &                  \\
\hline
    24    &     103    &      83.7323261    &      -5.3363055    &                         &  15.841$\pm$   0.016    &  13.369$\pm$   0.012    &  11.515$\pm$   0.007    &  10.718$\pm$   0.004    &  10.300$\pm$   0.030    &                        \\
    25    &            &      83.7545867    &      -5.3937160    &                         &                         &  14.382$\pm$   0.011    &  11.811$\pm$   0.005    &  11.018$\pm$   0.005    &  10.743$\pm$   0.017    &                        \\
    26    &     167    &      83.7570887    &      -5.4201451    &                         &                         &  15.427$\pm$   0.020    &  12.991$\pm$   0.014    &  12.173$\pm$   0.013    &  11.450$\pm$   0.036    &                        \\
    27    &     209    &      83.7672732    &      -5.3755507    &                         &  15.475$\pm$   0.013    &  13.013$\pm$   0.016    &  10.755$\pm$   0.004    &   9.768$\pm$   0.003    &   9.082$\pm$   0.007    &   8.151$\pm$   0.018   \\
    28    &     407    &      83.7957359    &      -5.2559454    &  16.583$\pm$   0.043    &  12.803$\pm$   0.037    &  10.836$\pm$   0.032    &   9.232$\pm$   0.001    &   8.464$\pm$   0.004    &   8.104$\pm$   0.004    &   7.707$\pm$   0.033   \\
    29    &     420    &      83.7970836    &      -5.4106294    &                         &  16.107$\pm$   0.023    &  12.677$\pm$   0.013    &   9.662$\pm$   0.008    &   8.171$\pm$   0.004    &   7.447$\pm$   0.014    &                        \\
    30    &     448    &      83.7993539    &      -5.3500809    &                         &  12.617$\pm$   0.024    &  10.011$\pm$   0.011    &   7.644$\pm$   0.002    &   6.807$\pm$   0.001    &   6.127$\pm$   0.005    &   5.326$\pm$   0.012   \\
    31    &     472    &      83.8007853    &      -5.4156826    &                         &  12.477$\pm$   0.009    &  10.698$\pm$   0.012    &   8.904$\pm$   0.003    &   8.157$\pm$   0.004    &   7.487$\pm$   0.013    &   6.636$\pm$   0.027   \\
    32    &            &      83.8058429    &      -5.3914514    &                         &                         &                         &   8.621$\pm$   0.009    &   7.010$\pm$   0.007    &   5.956$\pm$   0.008    &   5.096$\pm$   0.035   \\
    33    &     548    &      83.8064706    &      -5.2960711    &                         &  14.648$\pm$   0.009    &  11.908$\pm$   0.012    &   9.992$\pm$   0.002    &   9.181$\pm$   0.002    &   8.590$\pm$   0.010    &   7.577$\pm$   0.019   \\
    34    &     570    &      83.8071662    &      -5.3599724    &                         &  12.897$\pm$   0.023    &  10.624$\pm$   0.012    &   8.404$\pm$   0.011    &   7.445$\pm$   0.010    &   6.960$\pm$   0.032    &                        \\
    35    &            &      83.8074223    &      -5.3944417    &                         &                         &                         &   7.047$\pm$   0.005    &   5.170$\pm$   0.006    &   3.888$\pm$   0.004    &   2.994$\pm$   0.038   \\
    36    &     595    &      83.8079975    &      -5.4501972    &                         &  15.196$\pm$   0.014    &  11.901$\pm$   0.021    &   9.232$\pm$   0.004    &   8.236$\pm$   0.001    &   7.752$\pm$   0.012    &   7.369$\pm$   0.032   \\
    37    &            &      83.8195534    &      -5.3125430    &                         &                         &  13.387$\pm$   0.046    &   9.792$\pm$   0.008    &   8.530$\pm$   0.002    &   7.756$\pm$   0.018    &   7.093$\pm$   0.042   \\
    38    &    1030    &      83.8307100    &      -5.4283582    &  17.377$\pm$   0.083    &  13.356$\pm$   0.022    &  10.409$\pm$   0.016    &   7.654$\pm$   0.002    &   6.779$\pm$   0.001    &   6.034$\pm$   0.009    &   5.266$\pm$   0.021   \\
    39    &    1054    &      83.8325174    &      -5.2597844    &                         &  16.173$\pm$   0.059    &  12.783$\pm$   0.030    &   9.339$\pm$   0.002    &   8.100$\pm$   0.004    &   7.200$\pm$   0.002    &   6.411$\pm$   0.016   \\
    40    &    1094    &      83.8346095    &      -5.3326871    &                         &  13.958$\pm$   0.014    &  11.398$\pm$   0.008    &   8.927$\pm$   0.005    &   8.073$\pm$   0.004    &   7.824$\pm$   0.032    &                        \\
    41    &    1170    &      83.8408440    &      -5.3354363    &                         &  14.985$\pm$   0.016    &  12.799$\pm$   0.017    &  10.155$\pm$   0.006    &   9.281$\pm$   0.005    &   9.164$\pm$   0.110    &                        \\
    42    &    1188    &      83.8420775    &      -5.3160413    &                         &  16.641$\pm$   0.021    &  13.638$\pm$   0.014    &  10.825$\pm$   0.016    &   9.652$\pm$   0.006    &   8.930$\pm$   0.047    &                        \\
    43    &    1364    &      83.8604067    &      -5.3218452    &                         &  15.505$\pm$   0.018    &  12.678$\pm$   0.010    &  10.317$\pm$   0.008    &   9.364$\pm$   0.004    &   8.923$\pm$   0.032    &                        \\
    44    &    1480    &      83.8822676    &      -5.3556764    &                         &  14.409$\pm$   0.018    &  11.755$\pm$   0.016    &   9.401$\pm$   0.004    &   8.554$\pm$   0.002    &   8.868$\pm$   0.032    &                        \\
    45    &    1565    &      83.9236385    &      -5.3374070    &                         &                         &  15.703$\pm$   0.019    &  12.773$\pm$   0.013    &  11.751$\pm$   0.008    &  11.234$\pm$   0.035    &                        \\
\hline
\end{tabular}
\end{table}
\end{landscape}

%% file: tab7.tex
\begin{landscape}
\tabcolsep 0.06truecm
\begin{table}
\caption{NIR photometry of Class II candidates.}
\label{clII}
\centering
\begin{tabular}{ccccccccccc} 
\hline\hline
            Seq.   &            COUP   &        RA(2000)   &       Dec(2000)   &       J   &       H   &       K   &   [3.6]   &   [4.5]   &   [5.8]   &           [8.0]  \\
            Num.   &              ID   &           [deg]   &           [deg]   &           &           &           &           &           &           &                  \\
\hline
    46    &      12    &      83.6700479    &      -5.4440551    &  12.550$\pm$   0.024    &  11.652$\pm$   0.043    &  11.089$\pm$   0.029    &  10.416$\pm$   0.005    &  10.118$\pm$   0.006    &   9.754$\pm$   0.014    &   9.001$\pm$   0.016   \\
    47    &      13    &      83.6706135    &      -5.4332919    &  13.445$\pm$   0.018    &  12.312$\pm$   0.022    &  11.533$\pm$   0.019    &  10.571$\pm$   0.005    &  10.201$\pm$   0.006    &   9.904$\pm$   0.014    &   9.172$\pm$   0.017   \\
    48    &      21    &      83.6853649    &      -5.4106835    &  13.558$\pm$   0.013    &  12.778$\pm$   0.010    &  12.406$\pm$   0.013    &  11.811$\pm$   0.007    &  11.353$\pm$   0.010    &  11.119$\pm$   0.032    &  10.430$\pm$   0.107   \\
    49    &      29    &      83.6939237    &      -5.3903819    &  12.720$\pm$   0.013    &  11.771$\pm$   0.013    &  11.112$\pm$   0.013    &  10.240$\pm$   0.002    &   9.668$\pm$   0.004    &   9.113$\pm$   0.009    &   8.115$\pm$   0.027   \\
    50    &            &      83.6948357    &      -5.4346142    &  14.469$\pm$   0.024    &  12.810$\pm$   0.025    &  11.842$\pm$   0.023    &  10.443$\pm$   0.002    &   9.940$\pm$   0.002    &   9.608$\pm$   0.007    &   8.951$\pm$   0.011   \\
    51    &            &      83.6949818    &      -5.3580154    &  13.446$\pm$   0.013    &  12.466$\pm$   0.013    &  11.948$\pm$   0.012    &  11.151$\pm$   0.009    &  10.750$\pm$   0.007    &  10.291$\pm$   0.037    &   9.525$\pm$   0.060   \\
    52    &      37    &      83.6996888    &      -5.3944798    &                         &                         &                         &  12.039$\pm$   0.006    &  11.690$\pm$   0.007    &  11.278$\pm$   0.027    &  10.356$\pm$   0.046   \\
    53    &      40    &      83.7003928    &      -5.3772892    &  13.878$\pm$   0.013    &  13.138$\pm$   0.012    &  12.622$\pm$   0.010    &  12.052$\pm$   0.006    &  11.749$\pm$   0.010    &  11.367$\pm$   0.027    &  10.517$\pm$   0.079   \\
    54    &      44    &      83.7044071    &      -5.4406884    &  13.248$\pm$   0.026    &  12.420$\pm$   0.022    &  12.035$\pm$   0.017    &  11.302$\pm$   0.004    &  10.854$\pm$   0.004    &  10.541$\pm$   0.014    &   9.691$\pm$   0.022   \\
    55    &            &      83.7082042    &      -5.3123055    &   8.795$\pm$   0.041    &   8.383$\pm$   0.048    &   7.687$\pm$   0.085    &   6.022$\pm$   0.001    &   5.520$\pm$   0.001    &   5.129$\pm$   0.002    &   4.470$\pm$   0.001   \\
    56    &      54    &      83.7101265    &      -5.3389923    &  12.107$\pm$   0.030    &  11.183$\pm$   0.015    &  10.717$\pm$   0.016    &   9.871$\pm$   0.002    &   9.532$\pm$   0.003    &   9.279$\pm$   0.010    &   8.605$\pm$   0.023   \\
    57    &      55    &      83.7104375    &      -5.3930997    &  12.492$\pm$   0.024    &  11.626$\pm$   0.029    &  11.202$\pm$   0.019    &  10.872$\pm$   0.003    &  10.517$\pm$   0.004    &  10.190$\pm$   0.015    &   9.353$\pm$   0.029   \\
    58    &      58    &      83.7113833    &      -5.4502169    &  11.527$\pm$   0.030    &  10.750$\pm$   0.034    &  10.445$\pm$   0.022    &   9.557$\pm$   0.001    &   9.214$\pm$   0.002    &   8.926$\pm$   0.004    &   8.151$\pm$   0.006   \\
    59    &      62    &      83.7148246    &      -5.4201915    &                         &                         &   9.656$\pm$   0.020    &   8.562$\pm$   0.001    &   8.092$\pm$   0.001    &   7.725$\pm$   0.002    &   6.816$\pm$   0.002   \\
    60    &      66    &      83.7167237    &      -5.4119330    &  12.304$\pm$   0.037    &  11.256$\pm$   0.036    &  10.730$\pm$   0.035    &   9.748$\pm$   0.002    &   9.289$\pm$   0.001    &   8.944$\pm$   0.005    &   8.198$\pm$   0.006   \\
...\\
\hline
\end{tabular}
\end{table}
\end{landscape}

%% file: tab8.tex
\begin{landscape}
\tabcolsep 0.06truecm
\begin{table}
\caption{NIR photometry of Class III candidates.}
\label{clIII}
\centering
\begin{tabular}{ccccccccccc} 
\hline\hline
            Seq.   &            COUP   &        RA(2000)   &       Dec(2000)   &       J   &       H   &       K   &   [3.6]   &   [4.5]   &   [5.8]   &           [8.0]  \\
            Num.   &              ID   &           [deg]   &           [deg]   &           &           &           &           &           &           &                  \\
\hline
   194    &       1    &      83.6227896    &      -5.3936165    &  14.027$\pm$   0.011    &  13.316$\pm$   0.013    &  13.092$\pm$   0.011    &  12.757$\pm$   0.019    &  12.705$\pm$   0.024    &                         &                        \\
   195    &            &      83.6230879    &      -5.3954869    &  12.020$\pm$   0.029    &  10.659$\pm$   0.017    &  10.134$\pm$   0.015    &   9.780$\pm$   0.004    &   9.642$\pm$   0.004    &   9.625$\pm$   0.014    &   9.586$\pm$   0.026   \\
   196    &            &      83.6400806    &      -5.3860815    &  11.285$\pm$   0.027    &  10.754$\pm$   0.015    &  10.648$\pm$   0.014    &  10.743$\pm$   0.004    &  10.776$\pm$   0.007    &  10.794$\pm$   0.020    &  10.685$\pm$   0.069   \\
   197    &       6    &      83.6592044    &      -5.4065827    &  12.835$\pm$   0.012    &  11.990$\pm$   0.013    &  11.698$\pm$   0.017    &  11.544$\pm$   0.007    &  11.417$\pm$   0.010    &  11.525$\pm$   0.033    &                        \\
   198    &            &      83.6594147    &      -5.4007576    &  15.957$\pm$   0.014    &  14.627$\pm$   0.014    &  14.185$\pm$   0.017    &  14.127$\pm$   0.035    &  14.129$\pm$   0.062    &                         &                        \\
   199    &            &      83.6614465    &      -5.3569493    &  14.159$\pm$   0.016    &  13.023$\pm$   0.016    &  12.588$\pm$   0.021    &  12.409$\pm$   0.014    &  12.446$\pm$   0.020    &  12.528$\pm$   0.141    &                        \\
   200    &            &      83.6650038    &      -5.4217564    &  18.447$\pm$   0.044    &  16.860$\pm$   0.039    &  16.321$\pm$   0.023    &  15.897$\pm$   0.100    &  16.088$\pm$   0.182    &                         &                        \\
   201    &       7    &      83.6656279    &      -5.4071390    &   8.751$\pm$   0.034    &   8.408$\pm$   0.071    &   8.130$\pm$   0.079    &   7.890$\pm$   0.002    &   7.895$\pm$   0.003    &   7.838$\pm$   0.005    &   7.772$\pm$   0.007   \\
   202    &       9    &      83.6661159    &      -5.4449746    &  10.138$\pm$   0.025    &   9.908$\pm$   0.017    &   9.355$\pm$   0.016    &   9.417$\pm$   0.003    &   9.413$\pm$   0.004    &   9.400$\pm$   0.010    &   9.375$\pm$   0.019   \\
   203    &      10    &      83.6666439    &      -5.4343872    &  13.298$\pm$   0.018    &  12.589$\pm$   0.017    &  12.240$\pm$   0.015    &  11.951$\pm$   0.009    &  11.869$\pm$   0.013    &  11.863$\pm$   0.046    &                        \\
   204    &            &      83.6692738    &      -5.3876238    &  16.189$\pm$   0.013    &  15.246$\pm$   0.014    &  14.938$\pm$   0.027    &  14.723$\pm$   0.051    &  14.558$\pm$   0.069    &                         &                        \\
   205    &      14    &      83.6734241    &      -5.3992351    &  13.280$\pm$   0.012    &  12.558$\pm$   0.013    &  12.291$\pm$   0.013    &  11.952$\pm$   0.012    &  11.852$\pm$   0.016    &  11.862$\pm$   0.124    &                        \\
   206    &      16    &      83.6743312    &      -5.3637129    &  14.047$\pm$   0.013    &  13.394$\pm$   0.015    &  13.009$\pm$   0.013    &  12.587$\pm$   0.025    &  12.510$\pm$   0.023    &                         &                        \\
   207    &            &      83.6760625    &      -5.3758147    &  16.756$\pm$   0.017    &  15.252$\pm$   0.017    &  14.567$\pm$   0.019    &  14.280$\pm$   0.074    &  14.171$\pm$   0.101    &                         &                        \\
   208    &            &      83.6761015    &      -5.4214457    &  15.926$\pm$   0.018    &  14.885$\pm$   0.020    &  14.475$\pm$   0.019    &  14.164$\pm$   0.029    &  14.068$\pm$   0.041    &                         &                        \\
...\\
\hline
\end{tabular}
\end{table}
\end{landscape}

%% file: tab9.tex
\begin{table*}
\caption{X-ray photometry
 of the Class 0/I candidates taken from \citet{getm05a}.
 A description of the columns is given in Section\,\ref{xlf_section}.}
\label{coup_cl0I}
\centering
\begin{tabular}{cccccccccc} 
\hline\hline
            Seq.   &            COUP   &NetCts   &     PSF   &            Exp.   &       log $N_H$   &             kT1    &         log EM1    &   log $L_{h,c}$   &   log $L_{t,c}$  \\
            Num.   &              ID   &         &    Frac   &            (ks)   &     (cm$^{-2}$)   &           (keV)    &     (cm$^{-3}$)    & (erg\,s$^{-1}$)   & (erg\,s$^{-1}$)  \\
\hline
     3    &   274    &   176    &    0.86   &762.20    & 22.56$\pm$  0.06    & 15.00$\pm$ 15.00    & 52.28$\pm$  0.06    &           29.28   &           29.42  \\
     7    &   484    &    54    &    0.86   &783.40    & 22.54$\pm$  0.14    &  2.85$\pm$  1.67    & 51.96$\pm$  0.31    &           28.67   &           29.00  \\
     9    &   696    &    27    &    0.87   &825.80    & 22.16$\pm$  0.28    &  9.06$\pm$ 15.00    & 51.19$\pm$  0.31    &           28.19   &           28.38  \\
    10    &   702    &    79    &    0.86   &801.10    & 22.31$\pm$  0.17    &  3.25$\pm$  1.71    & 51.86$\pm$  0.27    &           28.63   &           28.93  \\
    11    &   860    &   185    &    0.87   &792.20    & 22.60$\pm$  0.12    &  5.73$\pm$  4.09    & 52.34$\pm$  0.20    &           29.27   &           29.49  \\
    12    &   859    &   481    &    0.87   &751.60    & 22.77$\pm$  0.05    &  1.90$\pm$  0.22    & 53.28$\pm$  0.12    &           29.78   &           30.24  \\
    16    &  1048    &   162    &    0.87   &795.80    & 22.68$\pm$  0.08    &  2.98$\pm$  0.84    & 52.51$\pm$  0.19    &           29.24   &           29.56  \\
    17    &  1115    &    27    &    0.86   &794.00    & 22.83$\pm$  0.19    &  1.13$\pm$  0.70    & 52.43$\pm$  1.00    &           28.56   &           29.37  \\
    19    &  1197    &    15    &    0.87   &799.30    &   ...               &   ...               &   ...               &          &  ...   \\
    21    &  1321    &   184    &    0.87   &714.40    & 23.11$\pm$  0.09    &  4.51$\pm$  2.98    & 52.82$\pm$  0.29    &           29.69   &           29.94  \\
    24    &   103    &    58    &    0.87   &783.40    & 22.63$\pm$  0.13    &  0.95$\pm$  0.35    & 52.64$\pm$  1.00    &           28.61   &           29.60  \\
    26    &   167    &    88    &    0.86   &739.20    & 22.73$\pm$  0.10    &  1.16$\pm$  0.24    & 52.79$\pm$  1.00    &           28.94   &           29.72  \\
    27    &   209    &    97    &    0.87   &824.10    & 22.37$\pm$  0.13    &  4.86$\pm$  3.60    & 51.90$\pm$  0.22    &           28.80   &           29.03  \\
    28    &   407    &   479    &    0.87   &737.40    & 22.63$\pm$  0.07    &  3.61$\pm$  0.99    & 52.91$\pm$  0.16    &           29.71   &           30.00  \\
    29    &   420    &    65    &    0.86   &838.20    & 22.29$\pm$  0.13    &  3.45$\pm$  2.65    & 51.70$\pm$  0.26    &           28.48   &           28.78  \\
    30    &   448    &  2249    &    0.86   &820.50    & 22.52$\pm$  0.02    &  2.24$\pm$  0.12    & 53.64$\pm$  0.05    &           30.24   &           30.64  \\
    31    &   472    &   501    &    0.86   &456.20    & 22.55$\pm$  0.05    &  2.80$\pm$  0.38    & 53.18$\pm$  0.10    &           29.88   &           30.21  \\
    33    &   548    &  2315    &    0.87   &774.50    & 22.54$\pm$  0.02    &  3.06$\pm$  0.25    & 53.59$\pm$  0.04    &           30.33   &           30.65  \\
    34    &   570    &  4070    &    0.87   &829.40    & 22.64$\pm$  0.02    &  3.35$\pm$  0.22    & 53.86$\pm$  0.03    &           30.63   &           30.93  \\
    36    &   595    &   624    &    0.87   &779.80    & 22.69$\pm$  0.04    &  2.31$\pm$  0.18    & 53.23$\pm$  0.10    &           29.84   &           30.23  \\
    38    &  1030    &  3616    &    0.87   &749.80    & 22.98$\pm$  0.02    & 15.00$\pm$  1.00    & 53.89$\pm$  0.03    &           30.88   &           31.01  \\
    39    &  1054    &  1630    &    0.88   &707.30    & 22.64$\pm$  0.02    &  2.59$\pm$  0.25    & 53.62$\pm$  0.07    &           30.28   &           30.64  \\
    40    &  1094    &   214    &    0.87   &801.10    & 22.61$\pm$  0.07    &  2.79$\pm$  0.55    & 52.60$\pm$  0.17    &           29.30   &           29.64  \\
    41    &  1170    &   150    &    0.87   &799.30    & 22.61$\pm$  0.08    &  1.03$\pm$  0.22    & 52.96$\pm$  1.00    &           29.02   &           29.91  \\
    42    &  1188    &   232    &    0.87   &686.10    & 22.73$\pm$  0.04    &  1.70$\pm$  0.27    & 53.10$\pm$  0.16    &           29.54   &           30.05  \\
    43    &  1364    &   501    &    0.88   &779.80    & 22.70$\pm$  0.03    &  4.30$\pm$  1.02    & 52.94$\pm$  0.11    &           29.80   &           30.05  \\
    44    &  1480    &   914    &    0.86   &684.40    & 22.69$\pm$  0.03    &  2.10$\pm$  0.18    & 53.51$\pm$  0.08    &           30.07   &           30.50  \\
    45    &  1565    &   385    &    0.88   &746.20    & 22.52$\pm$  0.05    &  2.50$\pm$  0.40    & 52.85$\pm$  0.12    &           29.50   &           29.86  \\
\hline
\end{tabular}
\end{table*}

%% file: tab10.tex
\tabcolsep 0.08truecm
\begin{landscape}
\begin{table*}
\caption{Upper Limits 
 of the X-ray undetected Class 0/I candidates.
 A description of the columns is given in Section\,\ref{xlf_section}.}
\label{cl0I_up_lim}
\centering
\begin{tabular}{cccccccccccc}
\hline\hline
            Seq.   &      RA\,(2000)   &     Dec\,(2000)   &             Cts   &             Cts   &            Exp.   & log $L_{h,c}^1$   & log $L_{h,c}^2$   & log $L_{h,c}^3$   & log $L_{t,c}^1$   & log $L_{t,c}^2$   & log $L_{t,c}^3$  \\
            Num.   &           (deg)   &           (deg)   &            hard   &            tot.   &           (ks)    & (erg\,s$^{-1}$)   & (erg\,s$^{-1}$)   & (erg\,s$^{-1}$)   & (erg\,s$^{-1}$)   & (erg\,s$^{-1}$)   & (erg\,s$^{-1}$)  \\
\hline
              1   &    83.70931200   &    -5.36415800   &           26.38   &           32.58   &          710.29   &           28.44   &           28.27   &           28.76   &           28.84   &           29.44   &           29.65  \\
              2   &    83.73136100   &    -5.29824800   &           69.32   &           86.29   &          739.17   &           28.84   &           28.67   &           29.16   &           29.25   &           29.85   &           30.05  \\
              4   &    83.79406000   &    -5.58508600   &          661.04   &         3605.31   &          681.67   &           29.85   &           29.69   &           30.18   &           30.90   &           31.50   &           31.71  \\
              5   &    83.79568709   &    -5.36277907   &           18.18   &           31.25   &          817.28   &           28.21   &           28.05   &           28.54   &           28.76   &           29.36   &           29.57  \\
              6   &    83.80009500   &    -5.31133400   &           21.00   &           26.16   &          781.40   &           28.29   &           28.13   &           28.62   &           28.70   &           29.30   &           29.51  \\
              8   &    83.81112700   &    -5.28122600   &           30.07   &           64.72   &          632.86   &           28.54   &           28.38   &           28.87   &           29.19   &           29.79   &           30.00  \\
             13   &    83.82558451   &    -5.34145849   &           16.28   &           21.07   &          800.83   &           28.17   &           28.01   &           28.50   &           28.60   &           29.20   &           29.41  \\
             14   &    83.83023800   &    -5.32007800   &           19.08   &           22.96   &          782.92   &           28.25   &           28.09   &           28.58   &           28.65   &           29.25   &           29.45  \\
             15   &    83.83193200   &    -5.29617200   &           27.37   &           31.85   &          764.63   &           28.42   &           28.25   &           28.74   &           28.80   &           29.40   &           29.61  \\
             18   &    83.83636500   &    -5.22322200   &         2873.07   &         4004.38   &          698.35   &           30.48   &           30.31   &           30.80   &           30.94   &           31.54   &           31.75  \\
             20   &    83.84798400   &    -5.46351500   &           19.52   &           23.76   &          784.75   &           28.26   &           28.09   &           28.58   &           28.66   &           29.26   &           29.47  \\
             22   &    83.86342600   &    -5.44756500   &           16.24   &           20.93   &          732.72   &           28.21   &           28.04   &           28.53   &           28.63   &           29.23   &           29.44  \\
             23   &    83.89975700   &    -5.35324100   &           19.93   &           23.70   &          651.91   &           28.35   &           28.18   &           28.67   &           28.74   &           29.34   &           29.55  \\
             25   &    83.75458500   &    -5.39371600   &           14.13   &           18.14   &          762.24   &           28.16   &           27.94   &           28.37   &           28.56   &           29.15   &           29.36  \\
             32   &    83.80584286   &    -5.39145145   &           14.85   &           19.82   &          346.70   &           28.52   &           28.30   &           28.73   &           28.94   &           29.54   &           29.74  \\
             35   &    83.80742228   &    -5.39444168   &           14.63   &           22.59   &          336.57   &           28.53   &           28.31   &           28.74   &           29.01   &           29.60   &           29.81  \\
             37   &    83.81955700   &    -5.31254300   &          846.89   &         1300.70   &          780.42   &           29.92   &           29.71   &           30.14   &           30.40   &           31.00   &           31.21  \\
\hline
\end{tabular}
\end{table*}
\end{landscape}

%% file: tab11.tex
\tabcolsep 0.1truecm
\begin{table*}
\caption{X-ray parameters
 of the Class II candidates. This table is available in its
 entirety in electronic form.}
\label{coup_clII_1}
\centering
\begin{tabular}{ccccccccc} 
\hline\hline
            Seq.   &            COUP    &       log $N_H$    &             kT1    &         log EM1    &             kT2    &         log EM2    &     log $L_{h,c}$   &   log $L_{t,c}$  \\
            Num.   &              ID    &     (cm$^{-2}$)    &           (keV)    &     (cm$^{-3}$)    &           (keV)    &     (cm$^{-3}$)    &   (erg\,s$^{-1}$)   & (erg\,s$^{-1}$)  \\
\hline
    46    &    12    & 21.33$\pm$  0.18    & 14.34$\pm$ 11.68    & 52.44$\pm$  0.05    &        ...          &        ...          &           29.48   &           29.64  \\
    47    &    13    & 20.00$\pm$  2.12    & 12.58$\pm$ 15.00    & 51.51$\pm$  0.17    &        ...          &        ...          &           28.54   &           28.71  \\
    48    &    21    & 20.79$\pm$  2.06    &  0.68$\pm$  0.11    & 51.63$\pm$  0.17    &  1.77$\pm$  0.47    & 51.89$\pm$  0.12    &           28.38   &           29.04  \\
    49    &    29    & 21.59$\pm$  0.03    &  0.81$\pm$  0.06    & 52.56$\pm$  0.13    &  3.13$\pm$  0.32    & 52.82$\pm$  0.03    &           29.59   &           30.04  \\
    52    &    37    & 20.00$\pm$  2.34    &  0.67$\pm$  0.33    & 51.13$\pm$  0.63    &        ...          &        ...          &           26.70   &           28.10  \\
    53    &    40    & 20.35$\pm$  2.00    &  0.86$\pm$  0.11    & 51.55$\pm$  0.15    &  6.29$\pm$  4.63    & 51.64$\pm$  0.08    &           28.62   &           28.99  \\
    54    &    44    & 21.79$\pm$  0.32    &  1.14$\pm$  0.35    & 51.50$\pm$  0.29    &        ...          &        ...          &           27.64   &           28.44  \\
...\\
\hline
\end{tabular}
\end{table*}

%% file: tab12.tex
\tabcolsep 0.1truecm
\begin{table*}
\caption{X-ray parameters
 of the Class III candidates. This table is available in its
 entirety in electronic form.}
\label{coup_clIII_1}
\centering
\begin{tabular}{ccccccccc} 
\hline\hline
            Seq.   &            COUP    &       log $N_H$    &             kT1    &         log EM1    &             kT2    &         log EM2    &   log $L_{h,c}$   &   log $L_{t,c}$  \\
            Num.   &              ID    &     (cm$^{-2}$)    &           (keV)    &     (cm$^{-3}$)    &           (keV)    &     (cm$^{-3}$)    & (erg\,s$^{-1}$)   & (erg\,s$^{-1}$)  \\
\hline
   194    &     6    & 20.67$\pm$  0.43    &  0.85$\pm$  0.12    & 52.17$\pm$  0.11    &  4.37$\pm$  0.95    & 52.62$\pm$  0.04    & 29.50   & 29.84  \\
   197    &     7    & 21.12$\pm$  0.12    &  0.78$\pm$  0.09    & 52.16$\pm$  0.12    &  2.32$\pm$  0.27    & 52.68$\pm$  0.04    & 29.32   & 29.79  \\
   201    &     9    & 20.93$\pm$  0.04    &  0.83$\pm$  0.01    & 53.58$\pm$  0.02    &  2.30$\pm$  0.08    & 53.94$\pm$  0.01    & 30.57   & 31.08  \\
   202    &    10    & 21.37$\pm$  0.03    &  0.83$\pm$  0.02    & 53.59$\pm$  0.03    &  3.04$\pm$  0.11    & 53.87$\pm$  0.01    & 30.64   & 31.08  \\
   203    &    14    & 20.00$\pm$  1.21    &  0.74$\pm$  0.13    & 51.42$\pm$  0.21    &        ...          &        ...          & 27.11   & 28.39  \\
   205    &    20    & 20.81$\pm$  0.14    &  0.69$\pm$  6.24    & 51.64$\pm$  1.00    &  9.23$\pm$ 15.00    & 51.19$\pm$  0.21    & 28.25   & 28.82  \\
...\\
\hline
\end{tabular}
\end{table*}

%% file: tab13.tex
\begin{landscape}
\tabcolsep 0.08truecm
\begin{table*}
\caption{Upper Limits 
 of the X-ray undetected Class\,II and III candidates}
\label{clII_III_up_lim}
\centering
\begin{tabular}{cccccccccccccc}
\hline\hline
            Seq.   &              ID   &      RA\,(2000)   &     Dec\,(2000)   &             Cts   &             Cts   &            Exp.   & log $L_{h,c}^1$   & log $L_{h,c}^2$   & log $L_{h,c}^3$   & log $L_{t,c}^1$   & log $L_{t,c}^2$   & log $L_{t,c}^3$   &           Class  \\
            Num.   &             H97   &           (deg)   &           (deg)   &            hard   &            tot.   &           (ks)    &  (erg\,s$^{-1}$   &  (erg\,s$^{-1}$   &  (erg\,s$^{-1}$   &  (erg\,s$^{-1}$   &  (erg\,s$^{-1}$   &  (erg\,s$^{-1}$   &                  \\
\hline
             51   &             85   &    83.69498179   &    -5.35801543   &             117   &             130   &            0.76   &           28.83   &           27.89   &           29.10   &           28.73   &           28.54   &           29.38   &              II  \\
             55   &            108   &    83.70820425   &    -5.31230548   &              83   &              66   &            0.45   &           28.92   &           27.99   &           29.19   &           28.66   &           28.47   &           29.31   &              II  \\
             66   &            179   &    83.74109068   &    -5.47828377   &              79   &             237   &            0.73   &           28.67   &           27.73   &           28.94   &           29.00   &           28.81   &           29.66   &              II  \\
             74   &            219   &    83.75862700   &    -5.30628963   &              27   &              35   &            0.77   &           28.19   &           27.26   &           28.46   &           28.15   &           27.96   &           28.81   &              II  \\
             98   &            294   &    83.77694276   &    -5.51927167   &              32   &              39   &            0.71   &           28.29   &           27.35   &           28.56   &           28.23   &           28.04   &           28.89   &              II  \\
            113   &            402   &    83.80437902   &    -5.56763586   &              57   &              66   &            0.66   &           28.58   &           27.64   &           28.85   &           28.50   &           28.31   &           29.15   &              II  \\
            123   &            449   &    83.81012490   &    -5.55521463   &              51   &             500   &            0.68   &           28.51   &           27.57   &           28.78   &           29.36   &           29.17   &           30.01   &              II  \\
            153   &            682   &    83.83838625   &    -5.55481843   &              47   &              57   &            0.71   &           28.46   &           27.52   &           28.73   &           28.40   &           28.21   &           29.05   &              II  \\
            166   &            789   &    83.85709585   &    -5.49312187   &              24   &              29   &            0.76   &           28.14   &           27.20   &           28.41   &           28.08   &           27.89   &           28.73   &              II  \\
            339   &            729   &    83.84592814   &    -5.49483049   &          23.07   &          28.00   &           0.72   &          28.12   &          27.21   &          28.41   &          27.92   &          27.85   &          28.77   &             III  \\
            210   &             62   &    83.67848933   &    -5.42114559   &          38.07   &          55.00   &           0.75   &          28.31   &          27.40   &          28.60   &          28.19   &          28.12   &          29.04   &             III  \\
\hline
\end{tabular}
\end{table*}
\end{landscape}

%% file: tab14.tex
\tabcolsep 0.15truecm
\begin{table*}
\caption{Literature Class\,0/I  
 counterparts.}
\label{classI_matches}
\centering
\begin{tabular}{ccccccccc}
\hline\hline
                Seq.   &               Class   &                COUP   &               LMH00   &              MLLA02   &               SBS04   &               SBS04   &               SBS05   &               RBP05  \\
                Num.   &                       &                       &                       &                       &                       &               name    &                       &                      \\
\hline
                   1   &              0/I\,a   &                       &                       &                       &                       &                       &                       &                      \\
                   2   &              0/I\,a   &                       &                       &                       &                       &                       &                       &                      \\
                   3   &              0/I\,a   &                 274   &                       &                       &                       &                       &                       &                      \\
                   4   &              0/I\,a   &                       &                       &                       &                       &                       &                       &                      \\
                   5   &              0/I\,a   &                       &    TPSC          63   &                 771   &                       &                       &                       &                      \\
                   6   &              0/I\,a   &                       &                       &                       &                       &                       &                       &                      \\
                   7   &              0/I\,a   &                 484   &    TPSC          50   &                       &                       &                       &                       &                      \\
                   8   &              0/I\,a   &                       &                       &                       &                       &                       &                       &                      \\
                   9   &              0/I\,a   &                 696   &    TPSC          59   &                       &                       &                       &                       &                      \\
                  10   &              0/I\,a   &                 702   &                       &                       &                       &                       &                       &                      \\
                  11   &              0/I\,a   &                 860   &                       &                       &                       &                       &                       &                      \\
                  12   &              0/I\,a   &                 859   &                       &                       &                       &                       &                       &                      \\
                  13   &              0/I\,a   &                       &    TPSC          56   &                 950   &                       &                       &                       &                      \\
                  14   &              0/I\,a   &                       &                       &                       &                       &                       &                       &                      \\
                  15   &              0/I\,a   &                       &                       &                       &                       &                       &                       &                      \\
                  16   &              0/I\,a   &                1048   &                       &                       &                       &                       &                       &                      \\
                  17   &              0/I\,a   &                1115   &                       &                       &                       &                       &                       &                      \\
                  18   &              0/I\,a   &                       &                       &                       &                       &                       &                       &                      \\
                  19   &              0/I\,a   &                1197   &                       &                 988   &                       &                       &                       &                      \\
                  20   &              0/I\,a   &                       &                       &                       &                       &                       &                       &                      \\
                  21   &              0/I\,a   &                1321   &                       &                 991   &                       &                       &                       &                      \\
                  22   &              0/I\,a   &                       &                       &                       &                       &                       &                       &                      \\
                  23   &              0/I\,a   &                       &                       &                       &                       &                       &                       &                      \\
                  24   &              0/I\,b   &                 103   &                       &                       &                       &                       &                       &                      \\
                  25   &              0/I\,b   &                       &                       &                       &                       &                       &                       &                      \\
                  26   &              0/I\,b   &                 167   &                       &                       &                       &                       &                       &                      \\
                  27   &              0/I\,b   &                 209   &                       &                       &                       &                       &                       &                      \\
                  28   &              0/I\,b   &                 407   &                       &                       &                       &                       &                       &                      \\
                  29   &              0/I\,b   &                 420   &    TPSC          74   &                       &                       &                       &                       &     MAX          18  \\
                  30   &              0/I\,b   &                 448   &    TPSC          65   &                       &                       &                       &                       &     MAX          22  \\
                  31   &              0/I\,b   &                 472   &    TPSC          31   &                       &                       &                       &                       &                      \\
                  32   &              0/I\,b   &                       &    TPSC           3   &                 369   &                  10   &             134-330   &                   7   &     MAX          40  \\
                  33   &              0/I\,b   &                 548   &                       &                       &                       &                       &                       &                      \\
                  34   &              0/I\,b   &                 570   &    TPSC          35   &                 797   &                       &                       &                       &     MAX          44  \\
                  35   &              0/I\,b   &                       &    TPSC          78   &                       &                   1   &             138-340   &                  16   &     MAX          46  \\
                  36   &              0/I\,b   &                 595   &                       &                       &                       &                       &                       &                      \\
                  37   &              0/I\,b   &                       &                       &                       &                       &                       &                       &                      \\
                  38   &              0/I\,b   &                1030   &    TPSC          41   &                       &                       &                       &                       &                      \\
                  39   &              0/I\,b   &                1054   &                       &                       &                       &                       &                       &                      \\
                  40   &              0/I\,b   &                1094   &    TPSC          73   &                1001   &                       &                       &                       &                      \\
                  41   &              0/I\,b   &                1170   &    TPSC          44   &                 990   &                       &                       &                       &                      \\
                  42   &              0/I\,b   &                1188   &                       &                       &                       &                       &                       &                      \\
                  43   &              0/I\,b   &                1364   &                       &                       &                       &                       &                       &                      \\
                  44   &              0/I\,b   &                1480   &                       &                       &                       &                       &                       &                      \\
                  45   &              0/I\,b   &                1565   &                       &                       &                       &                       &                       &                      \\
\hline
\end{tabular}
\end{table*}